\documentclass[ba,preprint]{imsart}

\RequirePackage{amsthm,amsmath,amsfonts,amssymb}
\RequirePackage[numbers]{natbib}
\RequirePackage[colorlinks,citecolor=blue,urlcolor=blue,backref=page,backref=page]{hyperref}
\RequirePackage{graphicx}
\usepackage{subcaption} 

\pubyear{2026}
\arxiv{	2505.22587 }
\volume{}
\issue{}
\firstpage{1}
\lastpage{1}

\makeatletter
\def\journal@name{}
\makeatother

\startlocaldefs
\theoremstyle{plain}

\newtheorem{theorem}{Theorem}[section]
\newtheorem{lemma}[theorem]{Lemma}
\newtheorem{proposition}[theorem]{Proposition} 
\newtheorem{corollary}[theorem]{Corollary}

\theoremstyle{definition}
\newtheorem{definition}[theorem]{Definition}

\theoremstyle{remark}
\newtheorem{remark}[theorem]{Remark}          

\endlocaldefs

\usepackage{tikz}
\usepackage{placeins}
\usepackage{hyperref}
\usepackage{algorithm}
\usepackage{algpseudocode}

\begin{document}

\begin{frontmatter}
\title{Bayesian Non-Parametric Inference for L\'evy Measures in State-Space Models}
\runtitle{BNP Inference for L\'evy Measures in SSM}

\begin{aug}
\author[A]{\fnms{Bill Z.}~\snm{Lin}\ead[label=e1]{zhl24@cam.ac.uk}},
\author[A]{\fnms{Simon}~\snm{Godsill}\ead[label=e2]{sjg30@cam.ac.uk}}

\address[A]{Department of Engineering, University of Cambridge\printead[presep={,\ }]{e1,e2}}

\runauthor{B.Z. Lin and S. Godsill}
\end{aug}


\maketitle

\begin{abstract}
L\'evy processes, known for their ability to model complex dynamics with skewness,  heavy tails, and discontinuities, play a critical role in stochastic modeling across various domains. However, inference for most L\'evy processes, whether in parametric or non-parametric settings, remains a significant challenge. In this work, we present a novel Bayesian non-parametric inference framework for inferring the L\'evy measures of subordinators and normal variance-mean (NVM) processes within a linear state space model. A flexible random measure, the Independent Gamma-scaled Dirichlet Process (IGSDP), is introduced, for which the well-known Gamma process is a special case,  leading to tractable conditional distributions for inference about both L\'evy measures. We further show that in the Gamma process special case, conjugacy can be achieved for hyper-parameter inference. An explicit characterization of the parameter contour for NVM processes is provided, enabling an identifiable parameterization of the model for effective Markov Chain Monte Carlo algorithms in posterior inference. The method is demonstrated on both synthetic and  tick-level (high-frequency) financial datasets.
\end{abstract}




\end{frontmatter}

\section{Introduction}
\subsection{Context and Motivation}
\label{Context and Motivation}
L\'evy processes, as a profound generalization of Brownian motion, accommodate both continuous and jump-based behavior, which allows them to model  more complex systems exhibiting skewness, heavy tails and discontinuities. These features make L\'evy processes particularly important in stochastic modeling across a wide range of fields. For example, in finance they play a crucial role in derivative pricing, with the Merton jump diffusion model being one of the earliest examples extending the classic Black-Scholes framework. In physics, L\'evy processes have been used extensively in fields such as quantum theory, continuous quantum measurements, and various other domains. Detailed reviews of these applications can be found in \cite{merton_jump_diffusion,tankov_book,levy_processes_theory_and_applications,time_changed_Levy_processes_theory,levy_momentum_trading,time_change_oxford_veraart2010time}. Moreover,  stochastic differential equations (SDEs) driven by L\'evy processes are a powerful extension, gaining  significant popularity in stochastic volatility modeling in finance, and a well-known example is the non-Gaussian Ornstein-Uhlenbeck-based model in \cite{levy_volatility_model}. More recently, the use of L\'evy-based SDEs has been studied in tracking applications, given their ability to capture more complex system dynamics, as demonstrated in works such as \cite{Gan_Ahmad_Godsill_2021,sharice_paper}.

Despite their versatility, the direct simulation and inference of L\'evy processes are generally challenging, with L\'evy-based SDEs posing further challenges that render many standard procedures inapplicable. These challenges arise primarily from the intractable likelihoods of most L\'evy processes and the infinite activity exhibited by many classes, such that an infinite number of jumps occur (almost surely) within any finite time interval. These make discrete path simulation and continuous process simulation based on jumps difficult. Hence, only a relatively limited subset of L\'evy processes is currently practical for application out of the vast range of theoretical possibilities, see \cite{tankov_book}.


A classical approach to simulation  uses random shot-noise series representations of L\'evy processes \cite{Rosinski2001,Khintchine1937}, which may be extended to the modeling of   L\'evy-driven stochastic differential equations (SDEs) \cite{samoradnitsky2017stable,Marcos}. Such simulation schemes open the door to sophisticated inference schemes using likelihood-based and Bayesian approaches \cite{Levy_State_Space_Model,sharice_paper,NGP_Dynamical_Model,Yaman_NGP}. These approaches can provide inference under limited data quantities and noisy data, providing (in the Bayesian case) regularization and uncertainty quantification to assist in decision-making processes, yielding meaningful results in scenarios where frequentist methods may struggle.

Building on these, we consider the setting where discrete, noisy, and potentially partial observations of low or moderate frequency are generated from a linear state-space model driven by the Normal Variance-Mean (NVM) mixture processes, a broad and applicable class of L\'{e}vy processes that is particularly amenable to computational inference \cite{More_Levy_Barndorff-Nielsen_Shephard,Basics_of_Levy_Barndorff-Nielsen_Shephard,NVM_Process_GH_Identifiability_Issue_mcneil2015quantitative,tankov_book}. We provide a methodology for direct inference about the underlying L\'{e}vy measures, which are generally unknown in real data, in addition to the other states and parameters of the system.

\subsection{Related Work}
\label{Related Work}


Prior work on non-parametric inference for L\'evy measures has primarily focused on directly observed L\'evy processes. In the frequentist setting, as summarized in \cite{frequentist_compound_poisson_approximation}, existing methods can be classified into three classes: projection methods \cite{figueroa-lopez_nonparametric_2004,figueroa2009nonparametric,frequentist_compound_poisson_approximation}, spectral methods \cite{neumann_nonparametric_2009,gugushvili2009nonparametric_spectral,gugushvili2012nonparametric_spectral,comte2009nonparametric_spectral,comte2011estimation_spectral}, and decompounding in the compound Poisson cases \cite{duval2013density_compound_poisson,Coca_2018,buchmann2003decompounding_compound_poisson}. Bayesian non-parametric inference for L\'evy measures is a relatively new and developing field. In the case of compound Poisson processes, \cite{Multidimensional_Decompounding} establishes the posterior contraction rate and consistency results in the multi-dimensional setting under regular discrete sampling; \cite{Decompounding} studies similar properties in the one-dimensional high-frequency setting. For practical implementations, \cite{Decompounding} develops an MCMC sampler after specializing to a finite Gaussian mixture specification, and \cite{gugushvili2020decompounding} develops a sampler for discrete jump distributions. \cite{Gamma_Type} studies Bayesian non-parametric inference for the L\'evy density within a structured class of Gamma-type subordinators. More recently, \cite{Gibbs_posterior_inference} provides a Gibbs posterior formulation of \cite{figueroa2009nonparametric,figueroa-lopez_nonparametric_2004}. However, all these methods rely on direct observations of L\'evy processes or their increments, and therefore do not directly apply to noisy and potentially partially observed state-space model settings.

The literature on non-parametric inference for L\'evy-driven SDEs and state-space models is more limited. \cite{belomestny2022nonparametric_gamma_driven_sde_volatility_estimate} studies the non-parametric Bayesian estimation for the volatility function in a Gamma-process-driven SDE. \cite{levy_driven_ou_non-parametric_2005} studies subordinator-driven scalar Ornstein–Uhlenbeck processes and estimates the L\'evy measure of the subordinator via spectral style methods by exploiting the self-decomposability of the stationary distribution, and hence is not applicable in our scenario. A recent theoretical contribution is \cite{BNP_state-dependent-Levy_Posterior_Consistency}, which establishes weak posterior consistency for Bayesian non-parametric inference of the drift and state-dependent jump coefficient in discretely observed jump diffusions with a unit diffusion coefficient. In contrast, our work studies Bayesian non-parametric inference of L\'evy measures for both the subordinator and NVM processes driving a linear state-space model under discrete, noisy, partial observations, requiring no stationarity assumptions and applicable to non-scalar models.


\subsection{Contributions}

In this work, we develop Bayesian  inference procedures for L\'evy measures within a linear L\'{e}vy-driven SDE framework \cite{Marcos,Levy_State_Space_Model}, presenting a Bayesian non-parametric approach that significantly extends the scope of earlier approaches. The models are complex and so inference is computed using an  augmented Markov Chain Monte Carlo (MCMC) algorithm, including inference for both the underlying L\'evy measures and the parameters/states of the SDE system. The proposed algorithms are tested on synthetic data generated from known L\'evy processes and also on high-frequency tick-level financial data, in order to demonstrate their accuracy and practical applicability.

We demonstrate, in particular, inference for L\'evy measures in the scenario of Normal Variance-Mean (NVM) L\'{e}vy processes driving linear SDEs, modeled as an independent subordinator L\'evy process driving a conditionally Gaussian model. This set-up is closely linked to time-changed models \cite{time_changed_Levy_processes_theory}, which have found many important applications, for example, in finance \cite{time_changed_levy_NIG_barndorff1997processes,time_changed_levy_NIG_SV_barndorff1997normal,time_changed_levy_VG_madan1990variance,time_changed_levy_VG_pricing_madan1998variance}. Special cases of the NVM processes are known to suffer from identifiability issues, such as the Generalized Hyperbolic processes \cite{GH_identifiability_alternative_solution_browne2015mixture,NVM_Process_GH_Identifiability_Issue_mcneil2015quantitative}. In this work, we provide a general formulation of the contour underlying the entire NVM process class and also treatments to achieve identifiability. Inference is achieved by posing the problem as a subordination mechanism in a linear state-space model, as introduced in \cite{Levy_State_Space_Model,Marcos}. We employ a flexible random measure, the Independent Gamma-scaled Dirichlet Process (IGSDP), as a prior on the subordinator L\'evy measure, which leads to an IGSDP mixture-driven SDE structure for the model. This allows us to infer the subordinator L\'evy measure via an IGSDP conditional posterior, and the NVM measure as an IGSDP mixture. The well-known Gamma process and its mixture are shown to be a special case of our framework, and new conjugacy results for hyper-parameter inference are presented for this case.

\subsection{Paper Structure}


The remainder of the paper is organized as follows: Section 2 gives the relevant preliminaries for the considered inference problem and framework. Section 3 introduces IGSDP, the hyper-parameter inference schemes, and the generative model. Section 4 details the parameter contour for the NVM process and the MCMC algorithm to realize the inference. Section 5 presents experimental results on synthetic and real financial data, and the practical utility of the inference results is demonstrated in terms of forecasting performance.

\section{Preliminaries}
\subsection{L\'evy Processes}


A L\'evy process $W=\{W(t):t\geq 0\}$ is a c\`adl\`ag (right-continuous with left limits) stochastic process, i.e.

\begin{equation}
    \label{Cadlag Definition for Process Delta}
    W(t-) = \lim_{s\to t,s<t} W(s),   \quad  \quad W(t)=W(t+) = \lim_{s\to t,s>t} W(s),
\end{equation} and at points of discontinuity, the jumps of the process are

\begin{equation}
    \label{Definition of Process Jumps}
    \Delta W(t) = W(t) - W(t-).
\end{equation} Furthermore, we have $W(0)=0$ almost surely, the increments are stationary and independent, and stochastic continuity applies \cite{tankov_book,Sato}. Its Characteristic Function (CF) is given by the L\'evy-Khintchine formula:

\begin{equation}
    \label{Levy Khintchine Formula}
        \mathbb{E}[e^{izW(t)}] = 
        \text{exp}\bigg\{ t \Big( -\frac{1}{2}\sigma^2 z^2 + ibz + \int_{\mathbb{R}\setminus\{0\}} \big(e^{izw} - 1 - izw\mathbb{I}_{|w| \leq 1} \big) \nu(dw) \Big) \bigg\}.
\end{equation} Such a representation is uniquely characterized by the L\'evy triplet $(b,\sigma^2,\nu)$, where $b\in \mathbb{R}$ is the drift term that accounts for both the deterministic drift $b_0$ and the expected contributions of small jumps

\begin{equation}
    \label{Levy Khintchine Rate Definition}
    b=b_0 + \int_{\mathbb{R}\setminus\{0\}}w\mathbb{I}_{|w|\leq 1}\nu(dw),
\end{equation}$\sigma^2 \in[0,\infty)$ is the variance coefficient of the Brownian motion component, and $\nu$ is the L\'evy measure, a positive Radon measure on $\mathbb{R}\setminus\{0\}$, defined as the measure for the expected number of jumps in some set $A \in \mathcal{B}(\mathbb{R})$ over a unit time interval:

\begin{equation}
    \label{Levy Measure Definition}
    \nu(A)=\mathbb{E}[\# \{t\in[0,1]:\Delta W_t \neq 0, \Delta W_t \in A\}]
\end{equation}
The drift component $b$ is finite for the class of L\'{e}vy processes considered in this paper, although not necessarily so in general L\'{e}vy processes.

The L\'evy measure characterizes the behavior of the pure-jump process component, satisfying the constraint: 

\begin{equation}
    \label{Integratability Constraint}
    \int_{\mathbb{R}} (1\wedge |w|^2)\nu(dw)<\infty.
\end{equation} The Radon-Nikodym derivative of the L\'evy measure with respect to the Lebesgue measure $Q(w)=\frac{\nu(dw)}{dw}$, when it exists, is commonly referred to as the L\'evy density. It can be interpreted as the Poisson arrival rate for jumps of size $w$. In the finite activity case where $\lambda = \int_{\mathbb{R}\setminus\{0\}}\nu(dw)<\infty$, the L\'evy measure has an intuitive interpretation via the decomposition:

\begin{equation}
    \label{Compound Poisson Decomposition of Levy Measure}
    \nu(dw) = \lambda f(dw),\quad \lambda = \int_{\mathbb{R}\setminus\{0\}}\nu(dw)<\infty,
\end{equation} where $\lambda$ is the overall Poisson rate parameter and $f(dw)$ is the jump size probability measure, which will here be assumed to possess a density function,  $f(dw)=f(w)dw$. The finite activity case is a  compound Poisson process and has a well-defined finite-activity series representation:
\begin{equation}
    \label{series representation for compound Poisson prcoesses}
    W(t) = \sum_{i=1}^{N_t} W_i \mathbb{I}(V_i \leq t),
\end{equation} where $N_t\sim Po(\lambda t)$ is the random Poisson number of jumps over the time interval $[0,t]$, $V_i$ is the $i$th ordered jump time, which can be thought of as the $i$th epoch of a uniform Poisson process with rate parameter $\lambda$, and $W_i$ is the size of the $i$th jump, drawn iid from $f(dw)$.

\subsection{Subordinator Processes}

An important sub-class of L\'evy processes is the class of non-decreasing, non-negative {\em subordinator processes\/}  $Z=\{Z(t):t\geq 0\}$, so-called since they are commonly employed in random time change of other L\'evy processes. From the L\'evy-Khintchine perspective \eqref{Levy Khintchine Formula}, a subordinator's characteristic triplet satisfies $\sigma^2=0$, $\nu((-\infty,0])=0$, $b\geq 0$, that is, $W(t)$ has no diffusion component, only positive jumps of finite variation and non-negative drift, satisfying the additional integrability condition \cite{tankov_book}:

\begin{equation}
    \label{subordinator process additional integratability condition}
    \int_0^\infty (w\wedge 1)\nu(dw)<\infty\,.
\end{equation} We consider the case of no additional constant drift $b_0$, so that the drift component \eqref{Levy Khintchine Rate Definition} arises entirely from small jumps and is finite by \eqref{subordinator process additional integratability condition}:

\begin{equation}
    \label{Pure jump subordinator drift}
    b=\int_0^\infty w\mathbb{I}_{|w|\leq 1}\nu(dw)<\infty.
\end{equation} This leads to the following simplified characteristic function:

\begin{equation}
    \label{Subordinator process characteristci function}
    \mathbb{E}[e^{izW(t)}] = 
        \text{exp}\bigg\{ t \Big( \int_0^\infty \big(e^{izw} - 1 \big) \nu(dw) \Big) \bigg\}.
\end{equation}

For the subordinator processes considered here, probability laws are often not known explicitly and may have infinite activity. 
Instead we rely on series representations of these processes, see 
 \cite{Khintchine1937,Rosinski2001}.
A series representation of subordinator processes takes a similar form to the general L\'{e}vy process above,
\begin{equation}
    \label{subordinator series representation}
    Z(t) = \sum_{i=1}^{\infty}Z_i \mathbb{I}(V_i\leq t),
\end{equation}  where $Z_i>0$ is the $i$th jump size and $V_i$ the corresponding jump time.
More general L\'evy processes may need additional centering terms in order to compensate for the accumulation of small jumps, see \cite{Rosinski2001,wolpert2021} for further details. For subordinators, however,
the stronger integrability condition \eqref{subordinator process additional integratability condition} implies that no centering is required. 
In many practical cases, the small jumps of infinite activity processes can be truncated without significant loss of accuracy \cite{tankov_book} provided that the small jump quadratic variation decays fast enough, so that an (almost surely) finite series approximation of \eqref{subordinator series representation} 
in the finite activity form of \eqref{series representation for compound Poisson prcoesses} may be obtained. If the small-jump quadratic variation decays more slowly than linearly with the truncation threshold, the residual process can be well approximated by a Brownian motion to improve the precision \cite{small_jump_process__Brownian_convergence_asmussen2001approximations,small_jump_SDE_Gaussian_convergence_fournier2011simulation,TVD_for_small_jump_process_convergence_into_Brownian_carpentier2021total,Marcos}. When the decay rate is too fast, as in the case of the Gamma process, a non-parametric density estimate for the small jump path can be used if an approximation is desired \cite{nonparametric_small_jump_process_path_density_estimation_duval2025nonparametric}.

\subsection{Normal Variance Mean (NVM) Processes}
\label{NVM Preliminary Section}

This study focuses on inference under the L\'evy state space framework proposed in \cite{Marcos,Levy_State_Space_Model}. The key driving L\'evy process for the model is a specific class of subordinated process, the Normal Variance-Mean (NVM) process $J(t)$ \cite{time_changed_Levy_processes_theory,Basics_of_Levy_Barndorff-Nielsen_Shephard,More_Levy_Barndorff-Nielsen_Shephard}, which is defined to be  Brownian motion $B(t)$ time-deformed by some subordinator process $Z(t)$,
\begin{equation}
    \label{NVM Process Definition}
    J(t) = \mu_w Z(t) + \sigma_w B(Z(t)),
\end{equation} where $\mu_w \in \mathbb{R}$ and $\sigma_w>0$ are the NVM subordination parameters. This specification leads to a broad family of processes with heavy tails and skewness, such as Variance Gamma, Normal Inverse Gaussian, and Generalized Hyperbolic processes. The L\'evy measure of the NVM process $\nu_{\text{NVM}}$ is related to that of the subordinator process $\nu$ by \cite{tankov_book,Rosinski2001},
\begin{equation}
    \label{NVM Levy Measure}
    \nu_{\text{NVM}}(dw) = \int_0^\infty N(dw;\mu_wz,\sigma_w^2 z) \nu(dz),
\end{equation}
 where $N(w ;\mu,\sigma^2)$ denotes the normal density with mean $\mu$ and variance $\sigma^2$.

Moreover, the series representation for the NVM process can be obtained by passing that of the subordinator process \eqref{subordinator series representation} through the subordination mechanism \eqref{NVM Process Definition}

\begin{equation}
    \label{Series representation for NVM Process}
    \begin{aligned}
        J(t) &= \sum_{i=1}^{\infty} J_i \mathbb{I}(V_i \leq t) =  \sum_{i=1}^{\infty}(\mu_wZ_i + \sigma_w \sqrt{Z_i}U_i) \mathbb{I}(V_i\leq t),
    \end{aligned}
\end{equation} where $U_i\overset{iid}{\sim} N(0,1)$ is a standard normal variable. Note that the NVM series shares the same jump times $\{V_i\}_{i\geq1}$ as the subordinator series, and therefore, the subordination mechanism in the series representation can be taken as a random modification of jump sizes.

\subsection{Shot Noise Representations of the L\'evy State Space Model}
\label{Shot Noise Representation Section}

Here, we review the key features and results for the shot noise representation of the L\'evy state space model \cite{Marcos,Levy_State_Space_Model}. The latent state $X(t)$ follows a linear SDE \cite{SDE}
\begin{equation}
    \label{Linear Latent State Dynamics}
    dX(t) = AX(t)dt + hdJ(t),
\end{equation} where $A$ is the transition matrix, $h$ is the noise mapping matrix, and $J(t)$ is the driving L\'evy process. We assume partial observations at discrete times $\{t_i\}$ from the following additive noise model:

\begin{equation}
    \label{Observation Noise Model}
    Y(t_i) = HX(t_i) + V(t_i),
\end{equation} with $H$ being the emission matrix and $V(t_i)$ being some observation noise. Starting from a point $s$ with $s\leq t$, the solution to \eqref{Linear Latent State Dynamics} is

\begin{equation}
    \label{Arbitrary Start Solution to Linear SDE}
    X(t) = e^{A(t-s)}X(s) + \int_s^t e^{A(t-u)}h dJ(u).
\end{equation} Substituting the series representation of L\'evy process into \eqref{Arbitrary Start Solution to Linear SDE} leads to the shot-noise representation of the system response. In the case of $J(t)$ being the NVM process \eqref{Series representation for NVM Process}, the representation is

\begin{equation}
    \label{Shot Noise Representation of System Response}
    \begin{aligned}
    &X(t) 
     = e^{A(t-s)}X(s) + \sum_{i=1}^{\infty} (\mu_w Z_i + \sigma_w \sqrt{Z_i}U_i) e^{A(t-V_i)}h \mathbb{I}(s < V_i \leq t).
    \end{aligned}
\end{equation} Detailed derivation and discussions for this representation are provided in \cite{Marcos,Rosinski2001}.

An important property of \eqref{Shot Noise Representation of System Response} is that, conditional on the subordinator series, \\$\{(Z_i,V_i)\}_{i\geq 1}$, $A$, and the NVM parameters $(\mu_w,\sigma_w^2)$, the response is Gaussian. If we assume Gaussian additive noise in \eqref{Observation Noise Model}, the Kalman filter and smoother \cite{KF_Reference} can be used to compute the likelihood for the observations and the posterior distributions of the latent states conditional on the subordinator series and the NVM parameters. Furthermore, the NVM parameters can be analytically marginalized by placing a Normal-Inverse-Gamma prior on $(\mu_w,\sigma_w^2)$, augmenting the latent state $X(t)$ with $\mu_w$, and defining the observation noise $V(t_i)\overset{iid}{\sim} N(0,\sigma_w^2 \bar{C}_v)$, with $\bar{C}_v$ being a fixed relative observation noise covariance. The observation likelihood conditional on the jump series $p(\{Y(t_j)\}_{j=1}^N|\{(Z_i,V_i)\}_{V_i \leq t_N})$ can then be computed as in \cite{Levy_State_Space_Model}; this assists in tractable inference in our problem, see Supplement A \cite{SuppA} for further  details. Where such a conjugate prior structure is deemed unsuitable, a slightly more complex inference procedure may be constructed that samples $\sigma_W^2$ as part of the MCMC.

\subsection{Dirichlet Process and Mixture}
\label{DP Process and Mixture Section}

The Dirichlet process (DP) is a stochastic process whose sample paths are probability measures that is commonly used as a prior distribution over the space of probability measures \cite{DP_source_Ferguson_1973}. \cite{Peter_Muller,Yee_Whye_Teh_DP} provide detailed reviews of the Dirichlet process. The DP is characterized by 2 parameters, the prior strength parameter $\alpha >0$, which is also referred to as the concentration parameter, and the base probability measure $H$ defined on $S$. A DP random measure $G$ is said to be distributed as:

\begin{equation}
    \label{DP Distirbution}
    G \sim \text{DP}(\alpha,H),
\end{equation} with $\mathbb{E}[G(\cdot)]=H(\cdot)$ and $\mathrm{Var}[G(\cdot)]=\frac{H(\cdot)(1-H(\cdot))}{1+\alpha}$.

Moreover, $G$ assigns probability $G(B)$ to every measurable set $B$ such that for each finite partition $\{B_1,...,B_k\}$ of $S$, the joint distribution of the random vector $(G(B_1),...,G(B_k))$ follows a Dirichlet distribution, which gives rise to its name:

\begin{equation}
    \label{DP marginal joint distribution over finite partition}
    (G(B_1),...,G(B_k)) \sim \text{Dir}(\alpha H(B_1),...,\alpha H(B_k)).
\end{equation}

In the setting of posterior inference, given a sequence of independent and identically distributed draws $\{\theta_j\}_{j=1}^n$ from the true underlying measure $G$, the DP posterior distribution for $G$ is:

\begin{equation}
    \label{DP Posterior Formula}
    G|\{\theta_j\}_{j=1}^n \sim \text{DP}\left(\alpha + n, \frac{\alpha}{\alpha+n}H + \frac{n}{\alpha+n}\sum_{j=1}^n \frac{1}{n}\delta_{\theta_j}\right),
\end{equation} where the posterior base is a mixture between the prior base and empirical distribution.

Another important property of the DP is its discrete nature. As a discrete random probability measure, $G$ can always be written as a weighted sum of point masses over the space:

\begin{equation}
    \label{DP as an infinite sum of point masses}
    G(\cdot) = \sum_{i=1}^\infty w_i \delta_{x_i}(\cdot),
\end{equation} with $w_i$ being the probability weights satisfying $\sum_{i=1}^\infty w_i = 1$ and $\delta_{x_i}$ being the Dirac measure at $x_i$. The well known Stick-breaking construction is utilised to realise this model \cite{stick_breaking_source_Sethuraman_1991}.

Inference for the DP is typically via MCMC \cite{neal2000markov,Ishwaran_James_2001} or variational inference \cite{VI_DPM_Blei_Jordan_2006}, and our focus here is on MCMC approaches. The infinite summation form in \eqref{DP as an infinite sum of point masses} is not practical for direct use, and there are two principal approaches for practical inference, as summarized in \cite{DP_Conditional_Marginal_Review_Favaro_Teh_2013}, the marginal and conditional samplers. The marginal samplers \cite{neal2000markov} marginalize out the random measure, while the conditional samplers \cite{slice_sampler_Kalli_Griffin_Walker_2011,Ishwaran_James_2001} apply random or deterministic truncation to \eqref{DP as an infinite sum of point masses}. See also \cite{CRM_Sampling_BA_zhang2025posterior} for a more detailed review of this topic. In the SSM setting, it is most convenient to use the conditional samplers because explicitly preserving the random measure induces a tractable conditional independence structure that simplifies inference under the complex dependency structure of our model. See Supplement A \cite{SuppA} for a finite DP sampler.

One important application of DP is in modeling mixtures, referred to as the Dirichlet process mixture (DPM), and it is defined as:

\begin{equation}
    \label{Dirichlet Process Mixture Model Definition}
    f_M(x) = \int K_{z}(x) dG(z),
\end{equation} where a discrete measure $G$ sampled from a DP is convolved with a continuous kernel $K_z$ to yield a continuous DPM density $f_M$.

Furthermore, the DPM model can be generalized into the mixture of DPM with respect to its hyper-parameters by sampling them. Assuming that the base measure is characterized by some parameters $\eta$ and $H_\eta$, the stick-breaking construction \eqref{DP as an infinite sum of point masses} leads to a $\eta$ likelihood for the unique draws $\{\theta^*_j\}_{j=1}^k$ from the DP, with $k\leq n$, where $n$ is the total number of draws from the true DP,

\begin{equation}
    \label{DP base measure parameter likelihood}
    p(\{\theta_j^*\}_{j=1}^k|\eta) = \prod_{j=1}^k H_{\eta}(\theta_j^*). 
\end{equation} For the DP model complexity parameter $\alpha$, its likelihood has been shown in \cite{DP_alpha_likelihood} to be:

\begin{equation}
    \label{DP alpha likelihood}
    p(k|\alpha,n) \propto \alpha^k \frac{\Gamma(\alpha)}{\Gamma(\alpha+n)}.
\end{equation} This likelihood allows hyper-parameter sampling by MH or, more effectively, the exact Gibbs step using an auxiliary variable \cite{DP_Alpha_Gibbs_Sampler,DP_hyperparameter_tutorial}. See Supplement A \cite{SuppA} for more details. This is known to provide model averaging and convergence benefits \cite{Bayesian_Model_Averaging_Hoeting_Madigan_Raftery_Volinsky,DP_Gaussian_Mixture_Model}

\section{Model Specification}
\label{Problem Formulation Session}

The aim of this work is  to perform Bayesian non-parametric inference of the L\'evy measure $\nu(dx)$, which is assumed to have the L\'evy density $Q(x) = \frac{\nu(dx)}{dx}$. Discrete observations $\{Y_j=Y(t_j)\}_{j=1}^N$ are assumed to be generated from the L\'evy state space model \eqref{Linear Latent State Dynamics} with noise \eqref{Observation Noise Model}, and sampled on an arbitrarily spaced, ordered time grid $\{t_j\}_{j=1}^N$, $t_{j+1}>t_j$. This model captures complex, non-Gaussian dynamics while retaining certain elements of tractability for simulation and inference. Moreover, its continuous-time formulation naturally accommodates irregularly spaced observations, making it directly applicable to challenging real-world problems such as tick-level financial data analysis or tracking problems. The L\'evy measures of the subordinator $Z(t)$ \eqref{subordinator series representation} and NVM L\'evy process $J(t)$ subordinated by it \eqref{NVM Process Definition} are inferred, and we introduce a new class of random measures to achieve this.

We begin this section with an introduction to the random measure and inference for its hyper-parameters, followed by the specification of the generative model.

\subsection{Independent Gamma-Scaled Dirichlet Process}

Here, we define and discuss the new class of random measures for the inference problem.

\begin{definition}[{\bfseries Independent Gamma-Scaled Dirichlet Process}]
\label{IGSDP Definition}

 Following the shape-rate convention, let $\lambda \sim \Gamma(\alpha_\lambda,\beta_\lambda)$, and independently, $G\sim \text{DP}(\alpha,H)$, where $H$ is a base probability measure defined on $S$. Define the random measure:
\begin{equation}
    \label{IGSDP Definition Equation}
    \hat{G} = \lambda G.
\end{equation}
We say that $\hat{G}$ follows an Independent Gamma-scaled Dirichlet Process (IGSDP):
\begin{equation}
    \hat{G} \sim \text{IGSDP}(\alpha_\lambda,\beta_\lambda,\alpha,H).
\end{equation}
    
\end{definition}

\begin{remark}
    \label{Stick-breaking construction of IGSDP Remark} By the stick-breaking construction of DP \eqref{DP as an infinite sum of point masses}, definition \ref{IGSDP Definition} implies a similar construction for IGSDP:
    \begin{equation}
    \label{IGSDP Stick-Breaking Representation}
        \hat{G}(\cdot) = \lambda \sum_{i=1}^\infty w_i \delta_{x_i}(\cdot) = \sum_{i=1}^\infty W_i \delta_{x_i}(\cdot),
    \end{equation} where $G(\cdot) = \sum_{i=1}^\infty w_i \delta_{x_i}(\cdot)$ is the stick-breaking construction of DP, and $W_i = \lambda w_i$. 
\end{remark}

\begin{remark}
\label{Remark: IGSDP Sampling}
    To sample from an IGSDP, we may draw $\lambda$ and $G$ independently, and their product is a valid sample of $\hat{G}$ from the IGSDP due to the deterministic mapping \eqref{IGSDP Definition Equation}. 
\end{remark} 

Furthermore, we note that the completely random measure, the Gamma process \cite{CRM_kingman1967}, is a special case of IGSDP. This is also the only completely random case for this process. See Supplement B \cite{SuppB} for more details.

\begin{proposition}
    \label{IGSDP Reduction to Gamma Process}
    Let $\hat{G} \sim \text{IGSDP}(\alpha_\lambda,\beta_\lambda,\alpha,H)$. If the shape and concentration parameters are set to be equal,  $\alpha_\lambda = \alpha$, then $\hat{G}$ becomes the Gamma process (GaP):
    \begin{equation}
        \hat{G} \sim \text{GaP}(\alpha,\beta_\lambda,H),
    \end{equation}i.e. given a fixed subset $A\subset S$, $\hat{G}(A)$ is Gamma distributed:
    \begin{equation}
    \label{Marginal Distribution of GaP Random Measure}
        \hat{G}(A) \sim \Gamma(\alpha H(A),\beta_\lambda).
    \end{equation}
\end{proposition}

\begin{proof}
    A DP random measure $G\sim \text{DP}(\alpha,H)$ is related to the GaP random measure $G'$ by the normalization \cite{DP_source_Ferguson_1973,CRM_Sampling_BA_zhang2025posterior}:

\begin{equation}
    \label{DP Relationship to GaP}
    G = \frac{G'}{\sum_{j=1}^K G'(A_j)},
\end{equation} where $\{A_j\}_{j=1}^K$ forms a disjoint partition of $S$. Using finite additivity and \eqref{Marginal Distribution of GaP Random Measure}, the normalization constant has the Gamma distribution:

\begin{equation}
    \label{Normalization constant distirbution of the Gamma Process}
    \lambda = \sum_{j=1}^KG'(A_j) = G'(S) \sim \Gamma(\alpha,\beta_\lambda). 
\end{equation} It is then straightforward to see that the Gamma process is a special case of the IGSDP according to definition \ref{IGSDP Definition}, and, in fact, the Gamma process is also referred to as the Gamma-scaled DP \cite{Gamma-scaled-DP-name-source}.
\end{proof}


\begin{remark}
    \label{Remark: IGSDP Conjugacy}
 Given the jump series $\{(Z_i,V_i)\}_{i=1}^M$ drawn from a L\'evy measure $\hat{G}$, IGSDP is the conjugate prior to $\hat{G}$. Let $\hat{G}\sim \text{IGSDP}(\alpha_\lambda,\beta_\lambda,\alpha,H)$. The prior equivalently introduces a DP prior to the jump size distribution $G$ and a Gamma prior to the rate parameter $\lambda$, as defined in \ref{IGSDP Definition}. Conditional independence is achieved for $G$ on $\{Z_i\}_{i=1}^M$ and $\lambda$ on $\{V_i\}_{i=1}^M$. The priors are conjugate in this case, leading to the posteriors:
    \begin{equation}
    \label{Posterior for the rate parameter}
    p(\lambda|\{V_i\}_{i=1}^M)=\Gamma(\alpha_\lambda+M,\beta_\lambda+\sum_{i=1}^MT_i),
\end{equation} where $\{T_i\}_{i=1}^M$ are the exponential samples converted from the consecutive differences of jump times with $V_0=0$,
\begin{equation}
    \label{Posterior on the jump size distribution}
        G | \{Z_i\}_{i=1}^M \sim \text{DP}\bigg(\alpha + M,\frac{\alpha}{\alpha + M} H_\eta + \frac{1}{\alpha + M} \sum_{i=1}^M \delta_{Z_i} \bigg).
\end{equation}  Then, by definition \ref{IGSDP Definition}, the posterior distribution is:

    \begin{equation}
        \label{Posterior IGSDP}
        \hat{G}|\{(Z_i,V_i)\}_{i=1}^M \sim \text{IGSDP}(\alpha_\lambda + M, \beta_\lambda + \sum_{i=1}^M T_i, \alpha+M,\frac{\alpha}{\alpha+M}H_\eta + \frac{1}{\alpha+M}\sum_{i=1}^M \delta_{Z_i}).
    \end{equation}
\end{remark}

\subsection{Hyper-Parameter Inference for IGSDP}

\label{Hyper-Parameter Sampling Section}
For general IGSDP priors defined in \ref{IGSDP Definition}, both $\eta$ and $\alpha$ likelihoods of the jump series are the same as the DP case in \eqref{DP base measure parameter likelihood} and \eqref{DP alpha likelihood}, and the same techniques can be applied by taking the latent components as the jump sizes. However, things are different in the Gamma process case, where we impose the constraint $\alpha_\lambda = \alpha$. Since $\alpha$ is now also the shape parameter for the rate, it has an additional likelihood from the jump times. This makes the Gamma prior a conjugate prior to $\alpha$.

\begin{proposition}
\label{Proposition for the Conjugate alpha inference for the Gamma process}
    Given jump series observations $\{(Z_i,V_i)\}_{i=1}^M$ drawn from a L\'evy measure with a Gamma process prior $\text{GaP}(\alpha,\beta,H)$, $p(\alpha)=\Gamma(\alpha ; a,b)$ is a conjugate prior, and the posterior is:
    \begin{equation}
     p(\alpha|\{(Z_i,V_i)\}_{i=1}^M) = \Gamma(\alpha;a+k,b-\log \frac{\beta}{\beta + \sum_{i=1}^MT_i}),
\end{equation} where $k$ is the number of unique jump sizes in the $M$ jumps and $T_i$ are the exponential samples as in \eqref{Posterior for the rate parameter}.
\end{proposition}

\begin{proof}
    We first note that the jump sizes and times are conditionally independent on $\alpha$:
    \begin{equation}
        \label{conditional independence on alpha for jump sizes and times}
        p(\{(Z_i,V_i)\}_{i=1}^M|\alpha) = p(\{Z_i\}_{i=1}^M|\alpha) p(\{V_i\}_{i=1}^M|\alpha). 
    \end{equation} Since exchangeability is preserved, the $\alpha$ likelihood for the jump sizes is the same as \eqref{DP alpha likelihood}. For the jump times, we marginalize over the $\lambda$ exponential likelihood:
    \begin{equation}
\label{alpha likelihood for jump times in GaP case}
    \begin{aligned}
        p(\{V_i\}_{i=1}^M|\alpha) & = \int p(\{V_i\}_{i=1}^M|\lambda) p(\lambda|\alpha) d\lambda \\
        &=\int \lambda^M e^{-\lambda \sum_{i=1}^MT_i} \times \frac{\beta^\alpha}{\Gamma(\alpha)}\lambda^{\alpha-1}e^{-\beta \lambda} d\lambda \\
        & = \frac{\beta^\alpha}{\Gamma(\alpha)} \frac{\Gamma(\alpha+M)}{(\beta + \sum_{i=1}^M T_i)^{\alpha + M}}.
    \end{aligned}
\end{equation} The posterior is then:

\begin{equation}
    \begin{aligned}
        p(\alpha|\{(Z_i,V_i)\}_{i=1}^M) &\propto p(\alpha)p(\{Z_i\}_{i=1}^M|\alpha)p(\{V_i\}_{i=1}^M|\alpha)\\
        & = \Gamma(\alpha;a,b) \times \alpha^k \frac{\Gamma(\alpha)}{\Gamma(\alpha+M)} \times \frac{\beta^\alpha}{\Gamma(\alpha)} \frac{\Gamma(\alpha+M)}{(\beta + \sum_{i=1}^M T_i)^{\alpha + M}} \\
        & \propto \alpha^{a+k-1} \exp\{-(b-\log \frac{\beta}{\beta+\sum_{i=1}^MT_i})\alpha\},
    \end{aligned}
\end{equation} which is a Gamma distribution:

\begin{equation}
     p(\alpha|\{(Z_i,V_i)\}_{i=1}^M) = \Gamma(\alpha;a+k,b-\log \frac{\beta}{\beta + \sum_{i=1}^MT_i}),
\end{equation} and note that $\log \frac{\beta}{\beta + \sum_{i=1}^MT_i}<0$, so the posterior rate parameter is always positive, and the distribution is valid. Proposition \ref{Proposition for the Conjugate alpha inference for the Gamma process} is therefore proven.

\end{proof}

This conjugate form for the IGSDP allows for closed-form conditional sampling of $\alpha$. 
However, we have found the additional flexibility of the general IGSDP with independent $\alpha$ and $\alpha_\lambda$ to give better performance in realistic applications. By construction, the jump size distribution and rate contribute independently to the jump series, so the additional dependency of the IGSDP imposes an unnecessary limitation. For instance, when we do not have much prior information and want small $\alpha$ values on the DP for weak prior strength, using the Gamma process prior leads to an unintended strong prior bias towards small values for the rate parameter.

\subsection{Generative Model}

Assigning an IGSDP prior to the subordinator L\'evy measure, the generative model for our problem may now be stated, using conjugate priors when available for the other model parameters, as shown below. 

We first generate the subordinator L\'evy measure $Q$ from an IGSDP prior with a concentration parameter $\alpha$, which is assigned a gamma prior,  from which the series of subordinator jumps $\{(Z_i,V_i)\}_{i\geq 1}$ can then be drawn. $(\mu_w,\sigma_w^2)$ are assigned a Normal-Inverse-Gamma prior distribution. Each NVM jump size is then assigned an independent Gaussian distribution $N(\mu_w Z_i, \sigma_w^2 Z_i)$. Finally, given the remaining model parameters $\theta$, the resulting NVM process drives the state space model to generate the observations $Y(t)$. This specification is detailed below:

\begin{equation}
\label{The Generative Model}
\begin{aligned}
\text{Subordinator $Z(t)$:}& \left\{
\begin{aligned}
     &\alpha \sim \Gamma(a,b)\\
    &Q|\alpha\sim \text{IGSDP}(\alpha_\lambda,\beta_\lambda,\alpha,H), \\
&(Z_i,V_i)|Q\sim Q,
\end{aligned}\right.
\\[2ex]
\text{NVM process $J(t)$:}& \left\{
\begin{aligned}
    & \sigma_w^2 \sim \text{IG}(\alpha_w, \beta_w), \\
    & \mu_w|\sigma_w^2 \sim N(\hat{\mu}_w,k_w \sigma_w^2),\\
    & J_i|Z_i, \mu_w, \sigma_w^2 \sim N(\mu_w Z_i, \sigma_w^2 Z_i),
\end{aligned}
\right. \\[2ex]
\text{State-space model $Y(t)$:}& \left\{
\begin{aligned}
    & \theta \sim p(\theta),\\
    & dX(t)= A_{\theta}X(t)dt + h_{\theta}dJ(t), \\
    & Y(t) = H_\theta X(t) + V(t), \quad V(t)\sim N(0,\sigma_w^2 \bar{C}_v). 
\end{aligned}
\right.
\end{aligned}
\end{equation}

Here the base measure $H$ must have positive support only, since it generates the L\'{e}vy measure for a subordinator $Z(t)$. Reasonable examples include the class of Generalized Inverse Gaussian \cite{GIG_Reference}, which includes the Inverse Gaussian, Gamma, and Inverse-Gamma distributions as special cases. For the observations $Y(t)$, we assume the system structure follows a parametric form with parameters $\theta$. These may be included in the MCMC inference procedure through an appropriate prior distribution $p(\theta)$.


Denoting the discrete observations  $\{Y_j\}_{j=1}^N$, where each $Y_j:=Y(t_j)$, and similarly for $X_j:=X(t_j)$, the target posterior distribution for inference in \eqref{The Generative Model} is then:
\begin{equation}
    \label{The full posterior distribution of the generative model}
    p(\alpha,Q,\{(Z_i,J_i,V_i)\}_{V_i\leq t_N},\mu_w,\sigma_w^2,\theta\,| \{Y_j\}_{j=1}^N).
\end{equation} Note that the system states $\{X_j\}_{j=1}^N$ or equivalently the NVM jumps $\{J_i\}_{V_i\leq t_N}$ are conditionally Gaussian given $\{(Z_i,V_i)\}_{V_i\leq t_N}$, as in \eqref{Series representation for NVM Process} and \eqref{Shot Noise Representation of System Response}. Furthermore, the Normal-Inverse-Gamma (NIG) prior for $(\mu_w,\sigma_w^2)$ leads to a conjugate posterior. Denote \\$\pi(\cdot)=p(\cdot|\{(Z_i,V_i)\}_{V_i\leq t_N},\theta,\{Y_j\}_{j=1}^N)$ to be distributions conditioned on the subordinator series, parameters and data,  then the posterior for $(\mu_w,\sigma_w^2)$ is:
\begin{equation}
    \label{pi posterior distirbution for muw}
    \pi(\mu_w|\sigma_w^2) = N(\mu_w;\mu',\sigma_w^2 k_w'),
\end{equation}
\begin{equation}
    \label{pi posterior distribution for sigmaw2}
    \pi(\sigma_w^2) = \text{IG}(\sigma_w^2;\alpha_w',\beta_w'),
\end{equation} with $\alpha'$ and $\beta'$ given in Supplement A \cite{SuppA}. These conditionally Gaussian and conjugate structures allow  $\{J_i\}_{i: V_i \leq t_N}$, and $(\mu_w,\sigma_w^2)$ to be analytically marginalized for inference, yielding the marginal posterior:
\begin{equation}
    \label{Marginal posterior distribution of the generative model}
     p(\alpha,Q,\{(Z_i,V_i)\}_{V_i\leq t_N},\theta|\{Y_j\}_{j=1}^N).
\end{equation}  A convenient conditional independence structure exploited in the MCMC  is \cite{graph_and_conditional_independence}

\begin{equation}
    \label{Conditional Independence in the Marginal Distribution}
    p(Q,\alpha|\{(Z_i,V_i)\}_{V_i\leq t_N},\theta,\{Y_j\}_{j=1}^N)=p(Q,\alpha|\{(Z_i,V_i)\}_{V_i\leq t_N}).
\end{equation}

The IGSDP prior for the subordinator L\'evy measure is, of course, of finite activity since, from its definition in \ref{IGSDP Definition}, the rate is Gamma distributed. We will also consider the use of our model in infinite activity cases via truncated series representations \cite{tankov_book} of \eqref{subordinator series representation}. In many practical settings, the quadratic variation decays sufficiently fast that the truncation introduces negligible error. Moreover, the small jumps are known to be difficult to identify, especially under the setting of noisy discrete observations over a finite time horizon with finite sampling frequency \cite{ssm_noise_inidentifiability_auger2016state,neumann_nonparametric_2009,figueroa2009nonparametric}. Therefore, it may be natural to consider performing inference while ignoring the residual small jumps in practice, even when the process is infinite-activity, as shown in later experiments.


\section{Posterior Inference}
\label{The MCMC Algorithm}
In this section, we describe the posterior inference procedure for the generative model \eqref{The Generative Model} under the marginal target posterior \eqref{Marginal posterior distribution of the generative model}. We begin by addressing the contour of the NVM parameters; then, the overall structure of the MCMC algorithm is presented, followed by the details of each sampling step and the posterior estimate for the NVM L\'evy measure.

\subsection{Contour of the NVM Parameters}
\label{Section: Contour of the NVM Parameters}
Here it is noted that the generative model in \eqref{The Generative Model} leads to  a contour of parameters corresponding to an equivalent NVM process, and hence an identifiability issue if the $Q(z)$, $\sigma_w$, and $\mu_w$ are to be jointly inferred. See the related discussion for the Generalized Hyperbolic processes  \cite{GH_original_barndorff1978hyperbolic} in \cite{NVM_Process_GH_Identifiability_Issue_mcneil2015quantitative,GH_identifiability_alternative_solution_browne2015mixture,review_GH_identifibaility_wei2019mixtures}. This may lead to difficulties in the convergence of the underlying parameters \cite{mcmc_contour_challenges_papamarkou2022challenges}. However, we note that if the goal is to infer just the NVM L\'{e}vy measure, it is not necessary to have a fully identifiable underlying model, and good mixing may still be achieved. See \cite{bayesian_neural_network_mixing_justifications_izmailov2021bayesian} for a relevant discussion. Now we propose a general formulation and treatment of the contour for the NVM processes. First,  in order to see the issue clearly, a lemma is provided,

\begin{lemma}
\label{contour lemma}
    For an NVM process $J(t)$ with parameters $(Q(z),\mu_w,\sigma_w^2)$, where $Q(z)$ is the L\'evy density of the subordinator process, there exists a contour of parameters having the  same law as $J(t)$:
    \begin{equation}
        \label{NVM Process Contour Definition}
        L = \{(\frac{1}{c}Q(\frac{1}{c}z),\frac{\mu_w}{c},\frac{\sigma_w^2}{c}):c > 0\}.
    \end{equation}
\end{lemma}

\begin{proof}
    This can be confirmed directly from the generative model for the NVM process:
    \begin{equation}
    \label{Identifiability Issue Equation of NVM Process}
    \begin{aligned}
        &J(t;Z(t),\mu_w,\sigma_w^2) = \mu_w Z(t) + \sigma_w B(Z(t))\\
        &= \frac{\mu_w}{c}cZ(t) + \frac{\sigma_w}{\sqrt{c}} B(cZ(t)) = J(t;cZ(t),\frac{\mu_w}{c},\frac{\sigma_w^2}{c}), \quad c>0,
    \end{aligned}
\end{equation} where $B(\cdot)$ is standard Brownian motion, and $c>0$ for a valid scaling of $\sigma_w^2$. Since  subordinator processes are closed under scaling, change of variable leads to the claim. Or, we may consider the corresponding NVM L\'evy density $\hat{Q}_{\text{NVM}}$ \eqref{NVM Levy Measure} for some value of $c$ on the contour $L$:

\begin{equation}
    \label{Equivalent NVM Levy Measure on the COntour Confirmation}
    \begin{aligned}
        &\hat{Q}_{\text{NVM}}(x) = \int_0^\infty N(x;\frac{\mu_w}{c}\hat{z},\frac{\sigma_w^2}{c}\hat{z})\frac{1}{|c|}Q_Z(\frac{1}{c}\hat{z}) d\hat{z}\\
        & = \int_0^\infty N(x;\mu_wz,\sigma_w^2 z) Q_Z(z) dz  = Q_{\text{NVM}}(x),
    \end{aligned}
\end{equation} which also verifies the claim. 
\end{proof}

In parametric modelling cases the identifiability issue is usually resolved by constraining the form and parameters of $Q$. Here though a simple solution is to fix the value of $\sigma_w^2$, which allows freedom of inference about $Q$
 while preventing the problem of an unidentified model, as in \cite{GH_identifiability_alternative_solution_browne2015mixture,NVM_Process_GH_Identifiability_Issue_mcneil2015quantitative}, which implies a fixed observation noise level $\sigma_v^2$.  

\subsection{The MCMC Sampler}

For the MCMC algorithm, we choose the primary inference target to be $Q$, whose posterior distribution is given by $p(Q|\{Y_j\}_{j=1}^N)$. As mentioned in section \ref{DP Process and Mixture Section}, a sampler similar to the conditional sampler structure \cite{Ishwaran_James_2001,slice_sampler_Kalli_Griffin_Walker_2011} is employed for inference in the SSM. Inference for the L\'{e}vy measure is then carried out by sampling from the augmented distribution $p(Q,\{(Z_i,V_i)\}_{V_i \leq t_N}|\{Y_j\}_{j=1}^N)$ in a collapsed Gibbs sampler \cite{collapsed_Gibbs_liu1994collapsed}, enabled by a Rao-Blackwellized structure that marginalizes $(\mu_W,\sigma_W^2)$ and $\{X(t)\}$ \cite{Levy_State_Space_Model}. See
\cite{Augmented_MCMC_Source,Augmented_MCMC_Comparison_Paper} for discussion of such generic schemes, noting that Metropolis-within-Gibbs \cite{Bishop_2006,Monte_Carlo_Statitsical_Methods_robert1999monte,MwG_Harris_Recurrence_roberts2006harris} and  Particle Gibbs \cite{particle_mcmc} are typical choices for the MCMC algorithms employed. Particle Gibbs has been found to be particularly powerful in inference for time series with strong temporal dependencies \cite{particle_gibbs_convergence,particle_gibbs_with_ancestor_sampling}, but with heavy computational cost, and here we opt instead for a simple and computationally cheap Metropolis-within-Gibbs approach, using overlapping-block updates \cite{Block_Updates} to further enhance its effectiveness. Other parameters may also be inferred by further augmentation of the target density, including the system parameters $\theta$ and the DP concentration parameter $\alpha$, targeting $p(Q,\alpha,\{(Z_i,V_i)\}_{V_i \leq t_N},\theta|\{Y_j\}_{j=1}^N)$. In principle the DP base measure hyperparameters can also be sampled but here we omit this owing to the additional challenges in MCMC mixing thereby introduced.

The algorithm can be summarized as: \begin{itemize}
    \item Initialize a subordinator series $\{(Z_i,V_i)\}_{V_i \leq t_N}^{(0)}$ and $\theta^{(0)},\alpha^{(0)}$ from appropriate distributions (such as their prior measures).

    \item Iterate between the following 3 steps. At iteration $k$, given the current state 
    $(Q^{(k-1)}, \alpha^{(k-1)}, \{(Z_i,V_i)\}^{(k-1)}_{V_i \le t_N}, \theta^{(k-1)})$, we update:
    \begin{enumerate}
        \item Sample the subordinator L\'evy measure $Q^{(k)}$ jointly with the DP concentration parameter $\alpha^{(k)}$ from $p(Q,\alpha|\{(Z_i,V_i)\}^{(k-1)}_{V_i \leq t_N},\theta^{(k-1)},\{Y_j\}_{j=1}^N)$.

        \item Sample $\{(Z_i,V_i)\}^{(k)}_{V_i\leq t_N}$ from $p(\{(Z_i,V_i)\}_{V_i\leq t_N}|Q^{(k)},\alpha^{(k)},\theta^{(k-1)},\{Y_j\}_{j=1}^N)$ 

        \item Sample $\theta^{(k)}$ from $p(\theta|\{(Z_i,V_i)\}^{(k)}_{V_i\leq t_N},Q^{(k)},\alpha^{(k)},\{Y_j\}_{j=1}^N)$
    \end{enumerate}
\end{itemize}
In the following sections, the details of how to conduct the 3 steps and perform inference for the NVM L\'evy measure are described.

\subsection{Sampling $Q$ and $\alpha$}

The sampling from this posterior is decomposed into 2 steps, utilising the conditional independence of the model \eqref{Conditional Independence in the Marginal Distribution}:

\begin{equation}
    \label{sampling by decompsoiton for the Levy measure and hyper-parameters}
    \begin{aligned}
        p(Q,\alpha&|\{(Z_i,V_i)\}_{V_i\leq t_N},\theta,\{Y_j\}_{j=1}^N)\\& = p(\alpha|\{(Z_i,V_i)\}_{V_i\leq t_N})\times p(Q|\alpha, \{(Z_i,V_i)\}_{V_i\leq t_N}),
    \end{aligned}
\end{equation} where the concentration parameter is sampled first, and $Q$ is then sampled conditionally, a partially collapsed scheme \cite{partially_collapsed_gibbs_van2008partially}.

For the first step, the posterior is reduced to $p(\alpha|\{Z_i\}_{i: V_i \leq t_N})$ for the unconstrained IGSDP, and the likelihood in \eqref{DP alpha likelihood} can be applied directly. In the Gamma process case, there is an additional condition on the jump times, and Proposition \ref{Proposition for the Conjugate alpha inference for the Gamma process} is used.

The second step, sampling from $p(Q|\{(Z_i,V_i)\}_{V_i \leq t_N},\alpha)$, is straightforward as mentioned in remarks \ref{Remark: IGSDP Sampling} and \ref{Remark: IGSDP Conjugacy}. This conditional sampler produces a finite IGSDP:

\begin{equation}
    \label{Discrete Levy Density Sample}
    Q^{(k)}(\cdot)= \lambda^{(k)}\sum_{j=1}^K w_j^{(k)}\delta_{z_j}^{(k)}(\cdot),
\end{equation} where $K$ is the truncated number of components.

\subsection{Sampling the Subordinator Series }

Conditioning on $Q$ makes $\{(Z_i,V_i)\}_{V_i \leq t_N}$ independent of the DP hyper-parameters, and to sample from the posterior we employ a Metropolis-within-Gibbs sampler. To further improve the sampling efficiency, an overlapping blocked update is applied, a well-established strategy in MCMC for state-space models \cite{Block_Updates}. The intuition is that sampling from the high-dimensional latent space via single-site Gibbs sampling can be computationally fast but with very slow convergence, while sampling the entire joint distribution in one step often leads to high rejection rates. The overlapping blocked updates offer a way to balance this trade-off. The block size and overlap are then treated as tuning parameters in the MCMC. For notational convenience, we denote $\{(Z_i,V_i)\}_{l:m} := \{(Z_i,V_i)\}_{V_i \in (t_{l-1},t_m]}$ for the jump series within the observation interval $(t_{l-1},t_m]$, and $\{(Z_i,V_i)\}_{-(l:m)} := \{(Z_i, V_i)\}_{V_i \le t_N} \setminus \{(Z_i, V_i)\}_{l:m}$.

 For each block update step, the following distribution is targeted with an MH step: 
\begin{equation}
    \label{single block update}
    p(\{(Z_i,V_i)\}_{l:m}|\{Y_j\}_{j=1}^N,Q,\{(Z_i,V_i)\}_{-(l:m)},\theta,\alpha).
\end{equation} For the proposal, we use $ p(\{(Z_i,V_i)\}_{l:m}|Q)$, which can be implemented using the simulation scheme provided in Supplement A \cite{SuppA}. The acceptance probability for each block update can be obtained in closed form due to the conditionally Gaussian structure described in section \ref{Shot Noise Representation Section}: 


\begin{equation}
    \label{acceptance probability in the MH within Gibbs step}
    \begin{aligned}
        &\alpha(\{(Z_i',V_i')\}_{l:m}|\{(Z_i,V_i)\}_{l:m})\\
        &= \text{min}\bigg(1, \frac{
        p(\{(Z_i,V_i)\}_{l:m}|Q)
        }{
        p(\{(Z'_i,V'_i)\}_{l:m}|Q)
        } 
        \frac{
        p(\{(Z'_i,V'_i)\}_{l:m}|\{Y_j\}_{j=1}^N,Q,\{(Z_i,V_i)\}_{-(l:m)},\theta) 
        }{
        p(\{(Z_i,V_i)\}_{l:m}|\{Y_j\}_{j=1}^N,Q,\{(Z_i,V_i)\}_{- (l:m) },\theta)
        } \bigg)\\
        & = \text{min}\bigg(1,\frac{
        p(\{Y_j\}_{j=1}^N|\{(Z'_i,V'_i)\}_{l:m},Q,\{(Z_i,V_i)\}_{- (l:m) },\theta)
        }{
        p(\{Y_j\}_{j=1}^N|\{(Z_i,V_i)\}_{l:m},Q,\{(Z_i,V_i)\}_{- (l:m) },\theta)
        } \bigg),
    \end{aligned}
\end{equation} which is the likelihood ratio for updating the section $(l:m)$ fraction of the jump series, and $\{(Z'_i,V'_i)\}_{l:m}$ is the proposed jump series.

\subsection{Sampling the System Parameters}

Sampling for $\theta$  uses a similar Metropolis-Hastings scheme. The acceptance probability for this step is: 

\begin{equation}
    \label{Langevin Parameter MH Acceptance Probability}
    \begin{aligned}
        &\alpha(\theta'|\theta)=\text{min}\bigg(1, \frac{q(\theta|\theta')p(\theta'|\{(Z_i,V_i)\}_{V_i\leq t_N},Q,\{Y_j\}_{j=1}^N)}{q(\theta'|\theta)p(\theta|\{(Z_i,V_i)\}_{V_i\leq t_N},Q,\{Y_j\}_{j=1}^N)}\bigg)\\
        & = \text{min}\bigg(1, \frac{q(\theta|\theta')p(\theta')p(\{Y_j\}_{j=1}^N|\{(Z_i,V_i)\}_{V_i\leq t_N},Q,\theta')}{q(\theta'|\theta)p(\theta)p(\{Y_j\}_{j=1}^N|\{(Z_i,V_i)\}_{V_i\leq t_N},Q,\theta)}\bigg),
    \end{aligned}
\end{equation} where $q(\theta'|\theta)$ is a suitable proposal function, and the likelihood for each $\theta$ value is computed as in  \eqref{acceptance probability in the MH within Gibbs step}.

\subsection{Posterior Mean Estimate of the NVM L\'evy Measure}

As seen in \eqref{NVM Levy Measure}  above, the subordination structure of the NVM model \eqref{NVM Process Definition} leads to an expression for the NVM L\'evy density through the convolution:

\begin{equation}
    \label{Deterministic Mapping for the NVM Measure}
    \begin{aligned}
        Q_{\text{NVM}}(x) &= 
        \int_0^\infty N(x;\mu_wz,\sigma_w^2 z) Q(dz):=\Phi(Q,\mu_w,\sigma_w^2).
    \end{aligned}
\end{equation} This deterministic mapping can be used for the construction of Monte Carlo-based sampling estimates of $Q_{\text{NVM}}$ based on posterior samples from $Q$. In our model, the conditionally Gaussian structure leads to a closed-form conditional posterior distribution for $\mu_w$ and $\sigma_w^2$, which allows for a Rao-Blackwellized posterior mean estimator for $Q_{\text{NVM}}$  through integration with respect to $\mu_w$ and $\sigma_w^2$.

\begin{proposition}
    \label{Proposition: NVM Levy Measure Conditional Estimate as Mixture of Student-t}
     Conditional on the jump series $\{(Z_i,V_i)\}_{V_i\leq t_N}$, the subordinator L\'{e}vy measure $Q$, model parameter $\theta$, and observations $\{Y_j\}_{j=1}^N$, the posterior mean of the NVM L\'evy measure under the priors in \eqref{The Generative Model} admits a Student-t kernel mixture representation
\begin{equation}
    \label{NVM Levy Measure Posterior Mean Estimate}
    \begin{aligned}
     &\mathbb{E}[Q_{\text{NVM}}(x)|Q,\{(Z_i,V_i)\}_{V_i \leq t_N},\theta,\{Y_j\}_{j=1}^N] \\
     & = \int_0^\infty St(x;\mu'z,\sqrt{\frac{\beta_w'(z + z^2k_w')}{\alpha_w'}},2\alpha_w') Q(dz),
    \end{aligned}
\end{equation} where the convention is location, scale, and degrees of freedom, and $\mu',k_w',\alpha_w'$ and $\beta_w'$ are the parameters in the NIG posterior for $(\mu_w,\sigma_w^2)$ in \eqref{pi posterior distirbution for muw} and \eqref{pi posterior distribution for sigmaw2}.

\begin{proof}

The posterior mean of the NVM L\'evy measure is obtained as:
\begin{equation}
\label{NVM Levy Measure Conditional Mean on Subordinator Measure}
\begin{aligned}
&\mathbb{E}[Q_{\text{NVM}}(x)| Q,\{(Z_i,V_i)\}_{V_i \leq t_N},\theta,\{Y_j\}_{j=1}^N]  
&= \iint \Phi(Q,\mu_w,\sigma_w^2)\; \pi(\mu_w,\sigma_w^2)\, d\mu_w\, d\sigma_w^2,
\end{aligned}
\end{equation} where $\Phi$ has been defined in \eqref{Deterministic Mapping for the NVM Measure}. Substituting \eqref{IGSDP Stick-Breaking Representation}, \eqref{final posterior distirbution for muw}, and \eqref{final posterior distribution for sigmaw2} into \eqref{NVM Levy Measure Conditional Mean on Subordinator Measure} leads to:

\begin{equation}
    \label{Mitxure of Student-t for IGSDP Mixture Estimate}
    \begin{aligned}
        &\mathbb{E}[Q_{\text{NVM}}(x)| Q,\{(Z_i,V_i)\}_{V_i \leq t_N},\theta,\{Y_j\}_{j=1}^N]\\
        & = \iint \int_0^\infty N(x;\mu_w z,\sigma_w^2 z)\, \; \pi(\mu_w,\sigma_w^2)\, Q(dz) \, d\mu_w\, d\sigma_w^2\\
        &= \int_0^\infty St(x;\mu'z,\sqrt{\frac{\beta_w'(z + z^2k_w')}{\alpha_w'}},2\alpha_w') Q(dz).
    \end{aligned}
\end{equation}

\end{proof}

\end{proposition}

\begin{corollary}
    If the subordinator L\'evy measure $Q$ admits the discrete form as in \eqref{IGSDP Stick-Breaking Representation}, the posterior mean becomes:
 \begin{equation}
    \label{NVM Levy Measure Posterior Mean Estimate for Discrete Subordinator Levy Denisty}
    \begin{aligned}
     &\mathbb{E}[Q_{\text{NVM}}(x)|Q,\{(Z_i,V_i)\}_{V_i \leq t_N},\theta,\{Y_j\}_{j=1}^N] \\
     &= \sum_{i=1}^\infty W_i St(x;\mu'z_i,\sqrt{\frac{\beta_w'(z_i + z_i^2k_w')}{\alpha_w'}},2\alpha_w'),
    \end{aligned}
\end{equation} where $\{z_i\}_{i=1}^\infty$ are the locations of the discrete measures in $Q$. To obtain the posterior mean conditional on just the observations, a Monte Carlo estimate may be used, which we will adopt later in the MCMC scheme:
\begin{equation}
    \label{MCMC Estimate for the Posterior Mean}
    \begin{aligned}
         &\mathbb{E}[Q_{\text{NVM}}(x)|\{Y_j\}_{j=1}^N] = \mathbb{E}[\mathbb{E}[Q_{\text{NVM}}(x)|Q,\{(Z_i,V_i)\}_{V_i \leq t_N},\theta,\{Y_j\}_{j=1}^N]|\{Y_j\}_{j=1}^N]\\
         & \approx \frac{1}{M} \sum_{m=1}^M\mathbb{E}[Q_{\text{NVM}}(x)|Q^{(m)},\{(Z_i^{(m)},V_i^{(m)})\}_{V_i^{(m)} \leq t_N},\theta^{(m)},\{Y_j\}_{j=1}^N],
    \end{aligned}
\end{equation} where $(Q^{(m)},\{(Z_i^{(m)},V_i^{(m)})\}_{V_i^{(m)}\le t_N},\theta^{(m)})\sim p(Q,\{(Z_i,V_i)\}_{V_i\le t_N},\theta\mid \{Y_j\}_{j=1}^N)$ are \\samples from the MCMC algorithm.
    
\end{corollary}

\section{Experiments}
\label{Section: Experiments}
In this section, the inference framework and algorithm are demonstrated on both synthetic and real data. The practical applicability of the method is demonstrated in terms of forecasting performance on real financial data.
\subsection{NVM Process Inference}
\label{Section: NVM Process Inference}
First consider the case of direct observations of the NVM process, which can be taken as a special case of our framework \eqref{The Generative Model} with a scalar $X(t)$, $A=0$, $h=1$, $H=1$, and $\bar{C}_v=0$. First, we demonstrate the performance of the algorithm in the case of an infinite activity subordinator, a tempered stable process $\text{TS}(\alpha,\beta,C)$, with $Q(z)=Cz^{-1-\alpha}e^{-\beta z}\mathbb{I}(z>0)$, where $C>0$, $\beta>0$, and $\alpha\in(0,1)$, and with subordinator jumps truncated at $10^{-7}$, using Algorithm 1 in \cite{Yaman_gig}, such that the truncated series has similar behavior to the exact infinite activity process, while retaining finite activity. Observations are generated from an NVM process having subordinator $\text{TS}(0.2,0.2,0.5)$, $\mu_w=1$, and $\sigma_w^2=1$, with $100$ data points uniformly spaced over $[0,10]$. Identifiability can be achieved by conditioning on $\sigma_w^2=1$, as discussed in Section \ref{Section: Contour of the NVM Parameters}. This choice allows for the recovery of the underlying NVM process parameters when such interpretability is desired, and we adopt it here to demonstrate the accuracy of the inference performance and also to assist in the diagnosis of the MCMC; however, we note that this constraint is unnecessary if the aim is to infer just the NVM measure.

\begin{figure}[H]
    \centering
    \begin{subfigure}{0.4\textwidth}
        \centering
        \includegraphics[width=\textwidth]{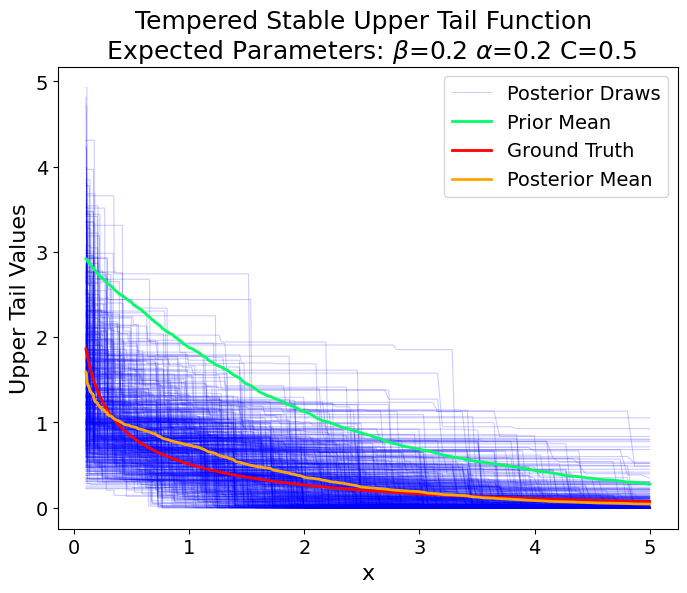}
        \caption{Subordinator L\'evy Measure}
        \label{nvm process case TS SUbordinator Measure Inferred}
    \end{subfigure}
    \hfill
    \begin{subfigure}{0.4\textwidth}
        \centering
        \includegraphics[width=\textwidth]{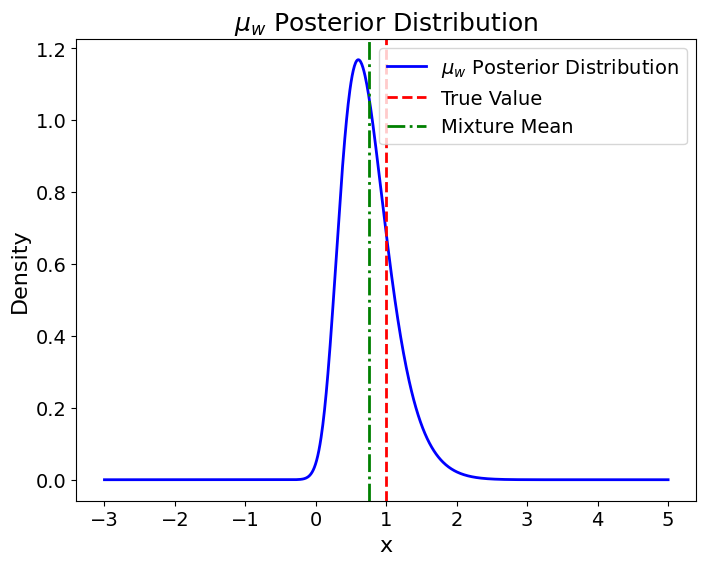} 
        \caption{  $\mu_w$ 
        }
        \label{nvm process muw inference}
    \end{subfigure}

    \begin{subfigure}{0.4\textwidth}
        \centering
        \includegraphics[width=\textwidth]{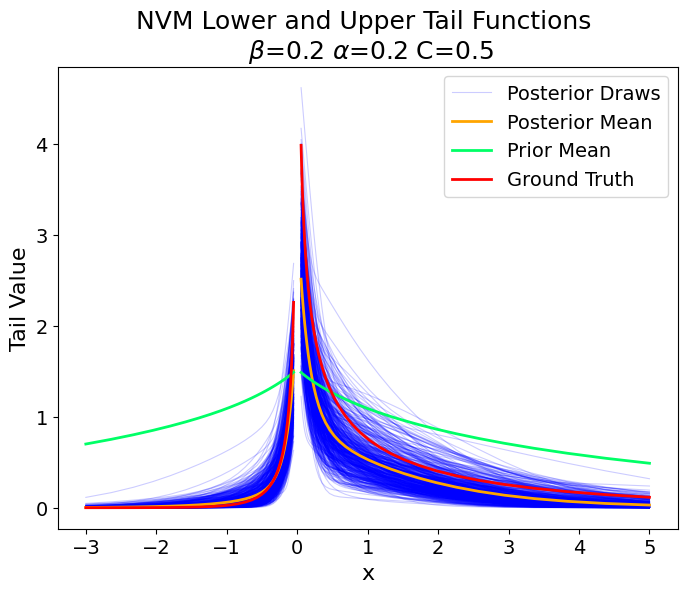}
        \caption{NVM L\'evy Measure}
        \label{nnvm process case vm measure inferred}
    \end{subfigure}
    \hfill
    \begin{subfigure}{0.5\textwidth}
        \centering
        \includegraphics[width=\textwidth]{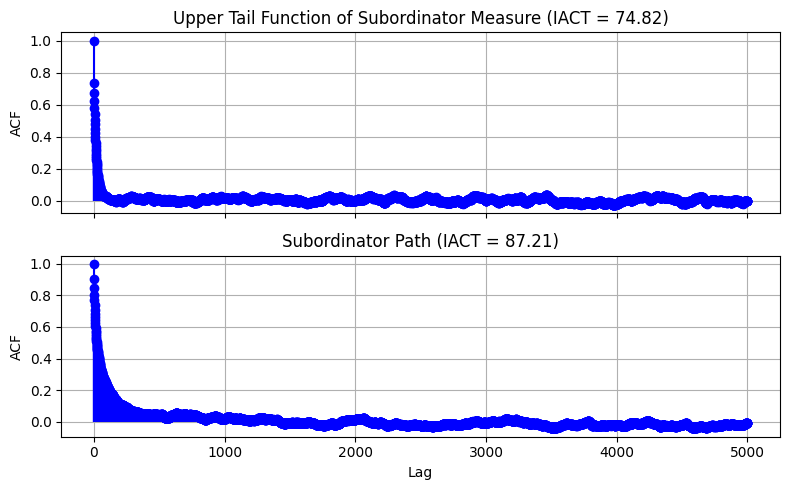}
        \caption{Mixing Performance}
    \label{nvm process case mixing performance}
    \end{subfigure}
    \caption{TS-NVM Process Inference Results}
    \label{TS-NVM Process Case Inference Results}
\end{figure}

Figure \ref{TS-NVM Process Case Inference Results} demonstrates the performance of the approach. The sub-block width and the overlap between sub-blocks in \eqref{acceptance probability in the MH within Gibbs step} have been chosen to be $5$ and $2$, respectively, and $150,000$ MCMC iterations were applied, discarding the first $50,000$ samples. Similar results were observed across many random initializations and repeated runs. For visualization, we consider the upper and lower tail functions for the L\'evy measures, $U(x)= \int_x^\infty \nu(du)$ and $L(x) = \int_{-\infty}^x \nu(du)$, as these are direct and stable functionals of the inferred random measures and avoid choice of smoothing kernel for the subordinator. For the prior model in \eqref{The Generative Model}, we chose priors that were quite different from the true L\'{e}vy measure and had large variances in order to demonstrate effective posterior inference. We put $\alpha \sim \Gamma(1,\frac{1}{3})$ following the shape-rate convention, $Q|\alpha \sim \text{IGSDP} (3,1,\alpha,\Gamma(1,\frac{1}{2}))$ as in Definition \ref{IGSDP Definition}, and $\mu_w\sim N(0,10)$. Figures \ref{nvm process case TS SUbordinator Measure Inferred} and \ref{nnvm process case vm measure inferred} show the inference results for the L\'evy measures, including the prior mean, posterior realizations, posterior mean, and the ground truth. Figure \ref{nvm process muw inference} shows the results for $\mu_w$, including the posterior distribution and the ground truth. We observe that despite the prior being deliberately misspecified and fairly diffuse, the posterior concentrates around the ground truths. Finally, inspired by \cite{Functional_ACT_Application}, we generalize the metrics in \cite{MCMC_in_statistical_mechanics} to compute the autocorrelation time for functions to analyze the mixing performance of the MCMC sampler, and the details are given in Supplement B \cite{SuppB}. Figure \ref{nvm process case mixing performance} shows the autocorrelation function (ACF) and the integrated autocorrelation time (IACT), noting that our burn-in time and number of iterations are considerably larger than the IACT.

In this simplest case, without any system structure, the frequentist estimators proposed in \cite{figueroa-lopez_nonparametric_2004,figueroa2009nonparametric} may be applied for comparison with our posterior mean estimate, and the implementation details can be found in Supplement B \cite{SuppB}. We employ the total variation norm, $\int_{\mathcal{X}} |Q_{\text{NVM}}(x)-\hat{Q}_{\text{NVM}}(x)|dx$, where $\mathcal{X}$ is the evaluation domain, to compare performance and also to tune the hyper-parameters in the frequentist methods. To further evaluate the performance of our approach, consider $2$ additional cases: the original NVM process driven by the tempered stable subordinator process under Gaussian observation noise with a standard deviation of $0.1$; and a NVM process driven by a bimodal subordinator process with a Poisson rate of $2$ and a bimodal jump size distribution, which is obtained via a mixture of Gamma distributions $0.7\Gamma(60,\frac{100}{3})+0.3 \Gamma(2000,200)$. In the bimodal case, a unimodal prior is deliberately placed on the subordinator L\'evy measure to demonstrate robustness to prior misspecification. We set $\alpha\sim\Gamma(0.5,2)$ for a weaker prior strength on the subordinator measure and $Q|\alpha \sim \text{IGSDP}(3,1,\alpha,\Gamma(1,\frac{1}{3}))$ as in Definition \ref{IGSDP Definition}, with a wider base measure to allow support over a broader region. This setting is referred to as bimodal NVM for brevity.

\begin{figure}[H]
    \centering

    \begin{subfigure}[H]{0.32\textwidth}
        \centering
        \includegraphics[width=\linewidth]{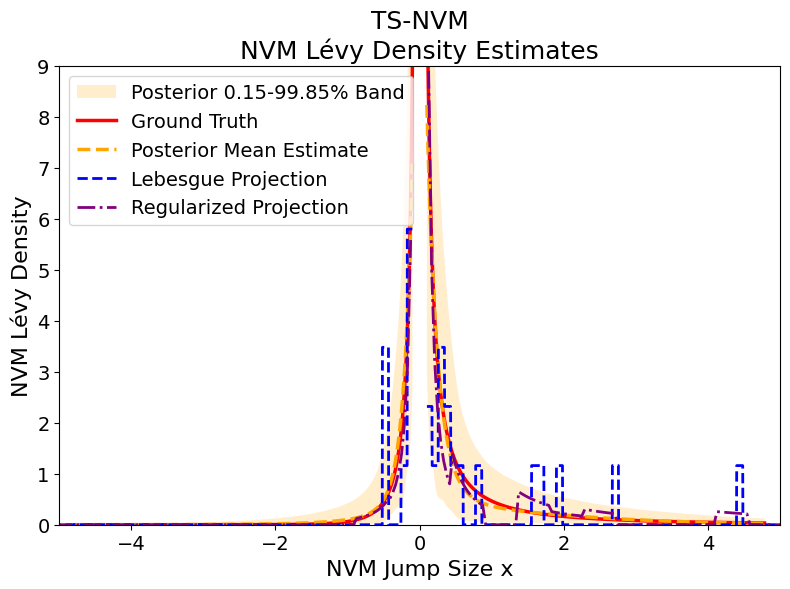}
        \caption{TS-NVM}
        \label{fig:ts_nvm_density}
    \end{subfigure}
    \hfill
    \begin{subfigure}[H]{0.32\textwidth}
        \centering
        \includegraphics[width=\linewidth]{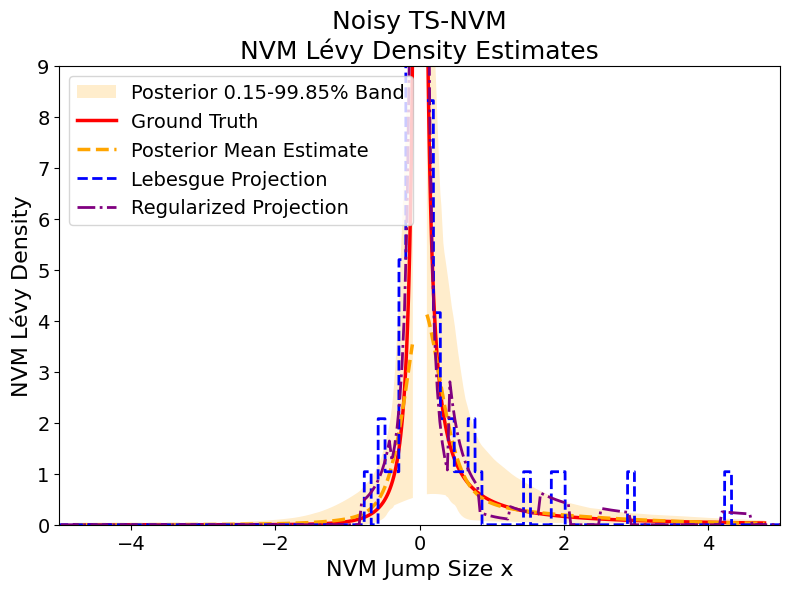}
        \caption{Noisy TS-NVM}
        \label{fig:noisy_ts_nvm_density}
    \end{subfigure}
    \hfill
    \begin{subfigure}[H]{0.32\textwidth}
        \centering
        \includegraphics[width=\linewidth]{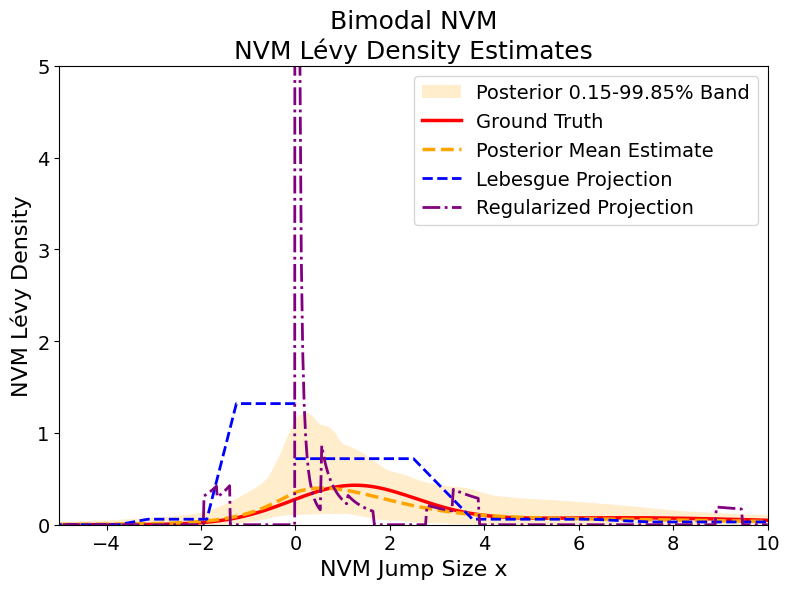}
        \caption{Bimodal NVM}
        \label{fig:bimodal_nvm_density}
    \end{subfigure}

    \caption{
    NVM L\'evy Density Estimate Comparison
    }
    \label{NVM L\'evy Density Estimate Comparison}
\end{figure}

\begin{figure}[H]
    \centering

    \begin{subfigure}[H]{0.32\textwidth}
        \centering
        \includegraphics[width=\linewidth]{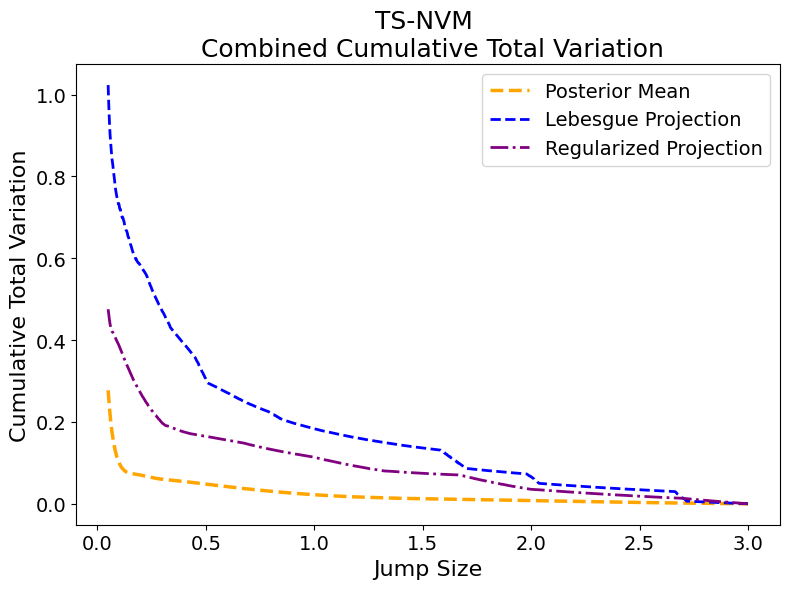}
        \caption{TS-NVM}
        \label{fig:ts_nvm_tvd}
    \end{subfigure}
    \hfill
    \begin{subfigure}[H]{0.32\textwidth}
        \centering
        \includegraphics[width=\linewidth]{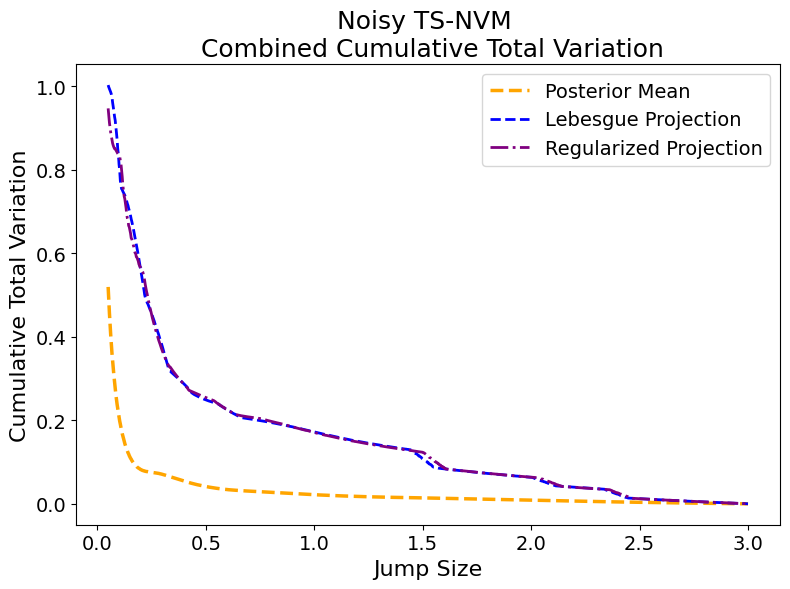}
        \caption{Noisy TS-NVM}
        \label{fig:noisy_ts_nvm_tvd}
    \end{subfigure}
    \hfill
    \begin{subfigure}[H]{0.32\textwidth}
        \centering
        \includegraphics[width=\linewidth]{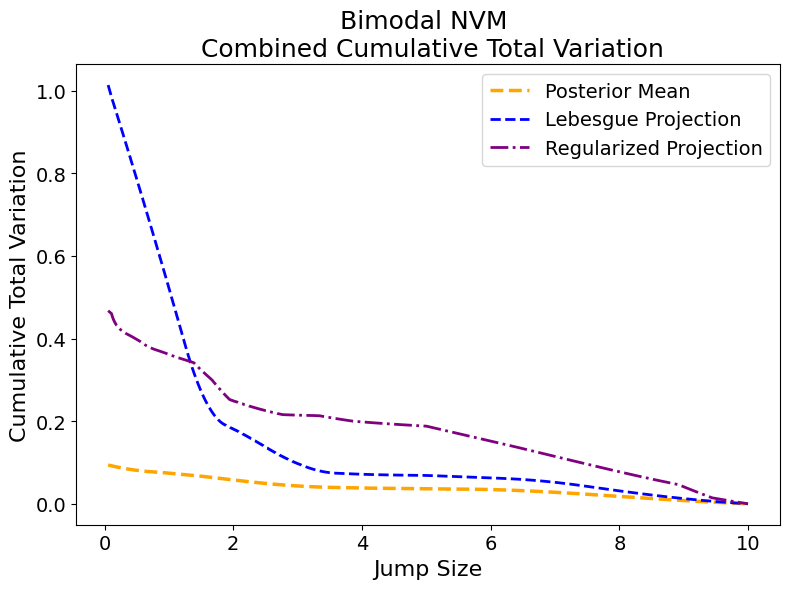}
        \caption{Bimodal NVM}
        \label{fig:bimodal_nvm_tvd}
    \end{subfigure}

    \caption{
   Cumulative Total Variation Norm Comparison
    }
    \label{Cumulative Total Variation Distance Comparison}
\end{figure}

Since the frequentist methods are density estimators, Figures \ref{NVM L\'evy Density Estimate Comparison} and \ref{Cumulative Total Variation Distance Comparison} compare the NVM density estimates and the corresponding cumulative total variation norms across the $3$ direct NVM process experiments. The experiments are designed to assess recovery accuracy in the zero-noise case, robustness to Gaussian observation noise, and robustness to prior misspecification. In addition to the point estimates, our method provides uncertainty quantification via the posterior band, which is useful in regions of weak identifiability. For the case without Gaussian observation noise, Figures \ref{fig:ts_nvm_density} and \ref{fig:ts_nvm_tvd} show that our estimator achieves superior performance in the density estimates. Under Gaussian observation noise, as shown in Figures \ref{fig:noisy_ts_nvm_density} and \ref{fig:noisy_ts_nvm_tvd}, while both frequentist estimators degrade in performance, especially the regularized estimator, our estimator is able to maintain comparably good estimates as in the case without noise, down to very small  jump sizes that are expected to be largely indistinguishable from the observation noise. The uncertainty band increases accordingly in the region of small jump sizes, indicating the reduced confidence in small jump estimation. Finally, Figures \ref{fig:bimodal_nvm_density} and \ref{fig:bimodal_nvm_tvd} show the performance in the bimodal case. Despite the misspecified prior, our posterior mean estimate matches well with the ground truth and achieves the best performance. For the regularized projection estimate, the mismatch in the regularizing measure leads to a strong bias that considerably degrades the quality of the estimate in this case.

\subsection{Langevin Model Inference}
\begin{figure}[H]
    \centering
    \begin{subfigure}{0.4\textwidth}
        \centering
        \includegraphics[width=\textwidth]{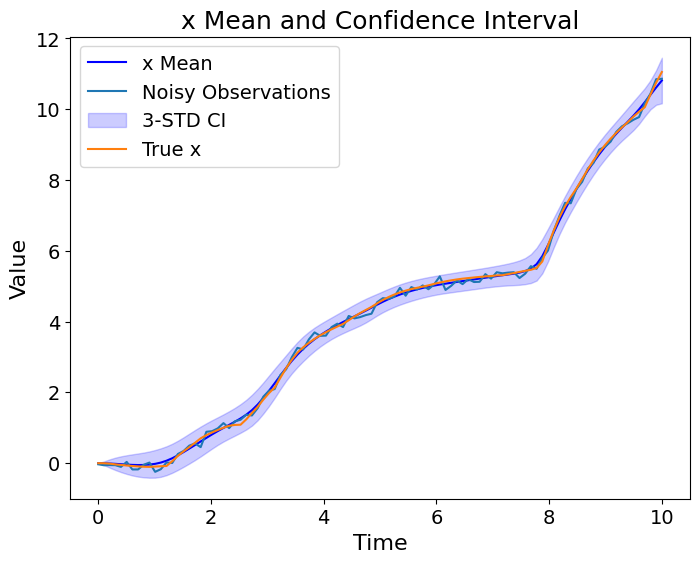}
        \caption{$x(t)$}
        \label{langevin x inference}
    \end{subfigure}
    \hfill
    \begin{subfigure}{0.4\textwidth}
        \centering
        \includegraphics[width=\textwidth]{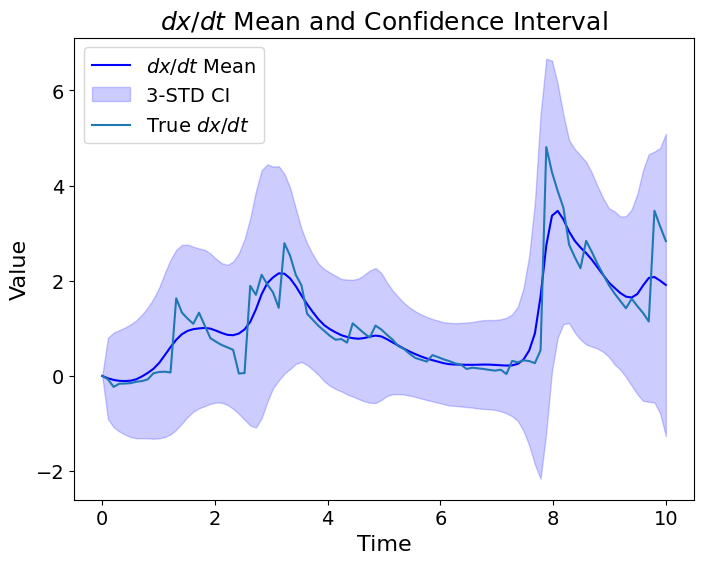} 
        \caption{$\frac{d}{dt}x(t)$}
        \label{langevin dxdt inference}
    \end{subfigure}

    \begin{subfigure}{0.4\textwidth}
        \centering
        \includegraphics[width=\textwidth]{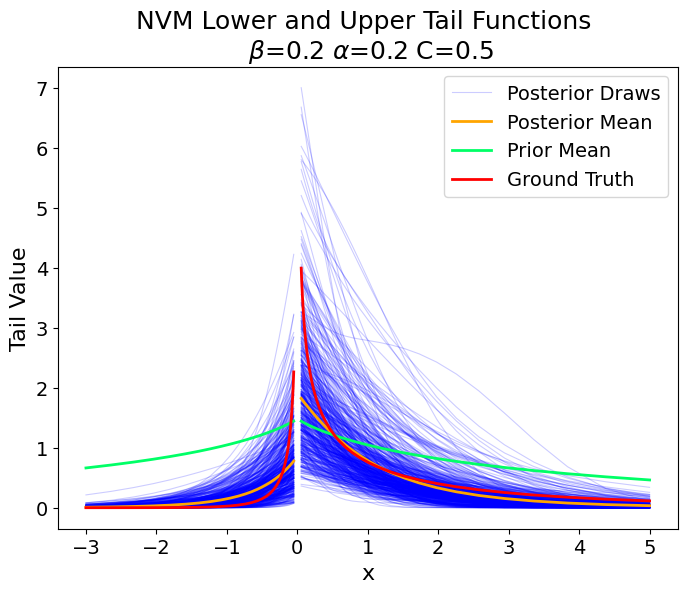}
        \caption{NVM L\'evy Measure}
        \label{langevin nvm levy measure inference}
    \end{subfigure}
    \hfill
    \begin{subfigure}{0.4\textwidth}
        \centering
        \includegraphics[width=\textwidth]{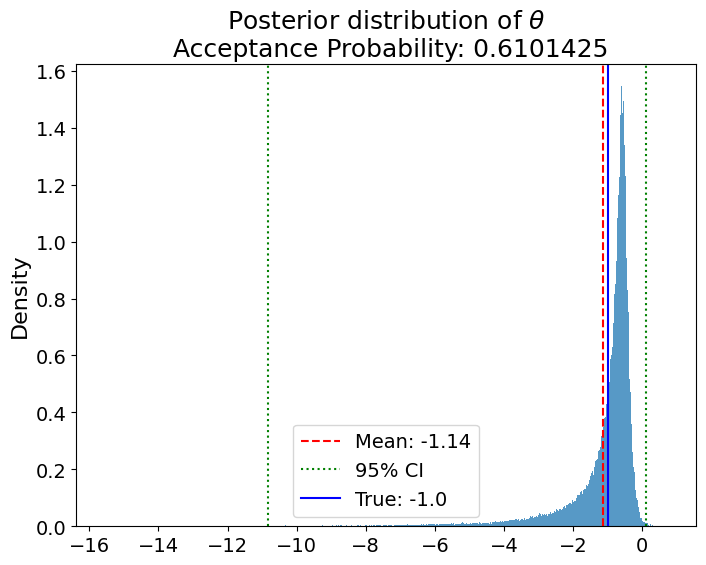}
        \caption{$\theta=-1$}
        \label{langevin theta inference}
    \end{subfigure}

    \caption{Inference Results for Langevin Model Driven by Tempered Stable Process }
    \label{Langevin Model Tempered Stable Process Inference Results}
\end{figure}

In this section, we demonstrate the power of our approach in dealing with complex state-space modeling scenarios. Here, the state-space model is a linear Langevin model driven by an NVM process and observed in Gaussian noise. We consider $2$ cases where the same NVM processes in Section \ref{Section: NVM Process Inference} are used to drive the Langevin model.

The Langevin model from physical, biological, and financial systems \cite{langevin_diffusion_reference_lemons1997paul,stochastic_processes_in_physics_and_chemistry_van1992stochastic} is defined within the general framework of \eqref{Linear Latent State Dynamics} and \eqref{Observation Noise Model}, with the system vectors/matrices defined as:
\begin{equation}
\label{Langevin System State Vector}
    X(t)
    = \begin{bmatrix}
        x(t) \\
        \frac{dx}{dt}(t) \\
    \end{bmatrix},\quad 
    A= \begin{bmatrix}
        0 & 1 \\
        0 & \theta \\
\end{bmatrix}, \quad 
    h= \begin{bmatrix}
        0 \\
        1 \\
\end{bmatrix}, \quad  H = \begin{bmatrix}
        1 & 0
    \end{bmatrix},
\end{equation}  so that the NVM process drives the velocity state, and $\theta$ is the mean reversion parameter. $\theta <0$ is required for a stable system. The choice of $H$ means that only the position $x(t)$ is observed under Gaussian noise. This provides a noisy and partially observed non-stationary state-space setting


\begin{figure}[H]
    \centering
    \begin{subfigure}{0.4\textwidth}
        \centering
        \includegraphics[width=\textwidth]{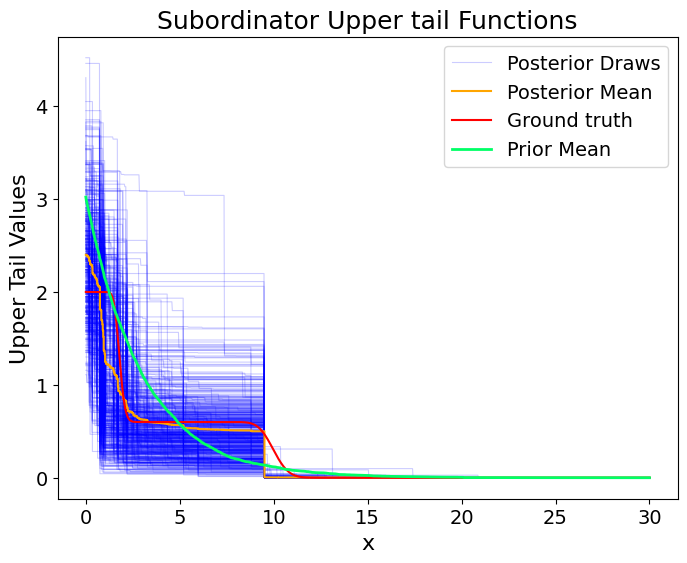}
        \caption{Subordinator L\'evy Measure}
        \label{bimodal case subordinator levy measure}
    \end{subfigure}
    \hfill
    \begin{subfigure}{0.4\textwidth}
        \centering
        \includegraphics[width=\textwidth]{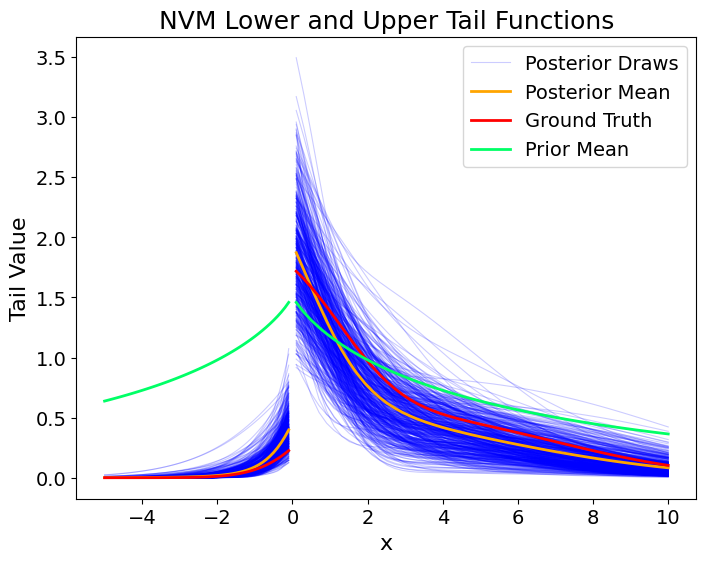} 
        \caption{NVM L\'evy Measure}
        \label{bimodal case nvm levy measure}
    \end{subfigure}
    \caption{Inference Results for Unimodal Prior on Bimodal L\'evy Measure}
\end{figure}

 Figure \ref{Langevin Model Tempered Stable Process Inference Results} shows the inference results for the case of the Langevin model \eqref{Langevin System State Vector} driven by the same NVM process with a tempered stable subordinator process as in Section \ref{Section: NVM Process Inference} and $\theta=-1$, under Gaussian observation noise with a standard deviation of $0.1$. The same prior model has been used, with an uninformative (flat) prior placed on $\theta<0$. The system states can be inferred via a collapsed Gibbs scheme \cite{collapsed_Gibbs_liu1994collapsed} by applying Monte Carlo marginalization of the conditionally Gaussian smoothing distributions given the subordinator series and system parameters, as in \eqref{Shot Noise Representation of System Response}, over the MCMC runs. See Supplement B \cite{SuppB} for more details. Figures \ref{langevin x inference} and \ref{langevin dxdt inference} show the inference results for the latent states in the Langevin model \eqref{Langevin System State Vector} with the posterior distribution and the ground truth. Figure \ref{langevin nvm levy measure inference} similarly shows the inference results for the NVM L\'evy measure. The posterior empirical distribution and ground truth for $\theta$ are shown in figure \ref{langevin theta inference}. In this more challenging state-space setting, both the system states and model structures are accurately inferred. In addition, the NVM L\'evy measure inferred also closely matches the ground truth down to small jump sizes, where some divergence may be expected because it is difficult to identify the small jumps in this challenging state-space model setting under Gaussian observation noise. It can also be argued that the inferred model is practically sufficient, given the accurate inference for the rest of the system.



Finally, we consider the same NVM process driven by a bimodal subordinator process under the same misspecified prior as in Section \ref{Section: NVM Process Inference}. Figure \ref{bimodal case subordinator levy measure} shows the unimodal prior along with the inferred measure and posterior realizations. Figure \ref{bimodal case nvm levy measure} shows the inference performance for the NVM L\'evy measure in this case. Despite the misspecified prior, the posteriors converge closely to the ground truths, indicating the ability of the method to infer unusual jump distributions.

\subsection{Application to High-Frequency Tick-Level Financial Data}

Finally, having demonstrated the effectiveness of the algorithm on synthetic datasets, the method is now applied to tick-level foreign exchange (FX) data with an irregularly spaced time axis. We use the GBP/USD data in January 2025 from TrueFX \cite{TrueFX_GBPUSD_2025_01}. The data is scaled based on the order of magnitude of the variation, and the initial subordinator series is generated from a tempered stable process with parameters chosen to yield comparable magnitudes. Training is done on $100$ ticks of the data, and details for the training setting and results can be found in Supplement B \cite{SuppB}.

\begin{figure}[H]
    \centering
    \begin{subfigure}[b]{0.7\linewidth}
        \centering
        \includegraphics[width=\linewidth]{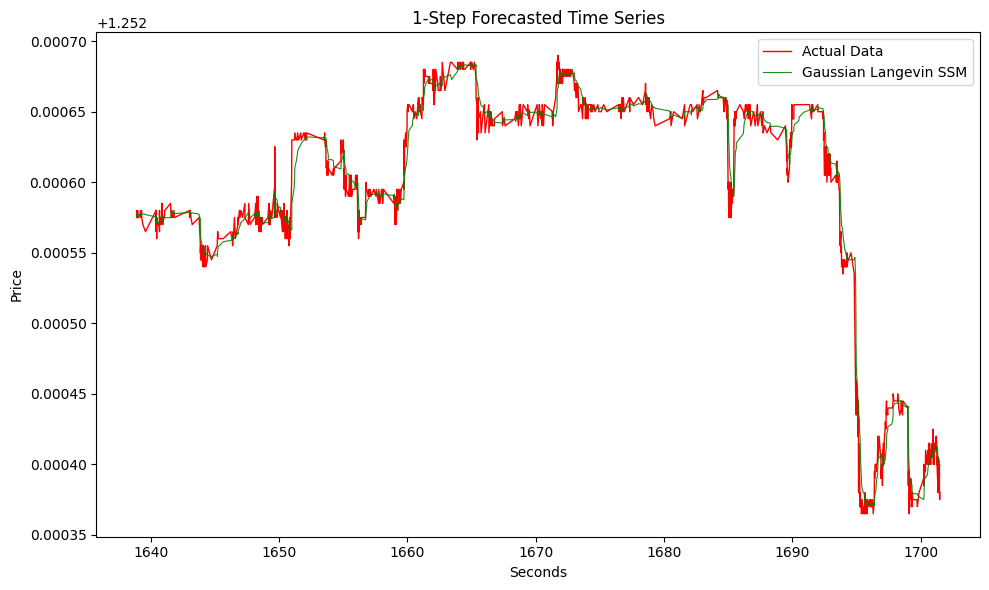}

    \end{subfigure}

    \begin{subfigure}[b]{0.7\linewidth}
        \centering
        \includegraphics[width=\linewidth]{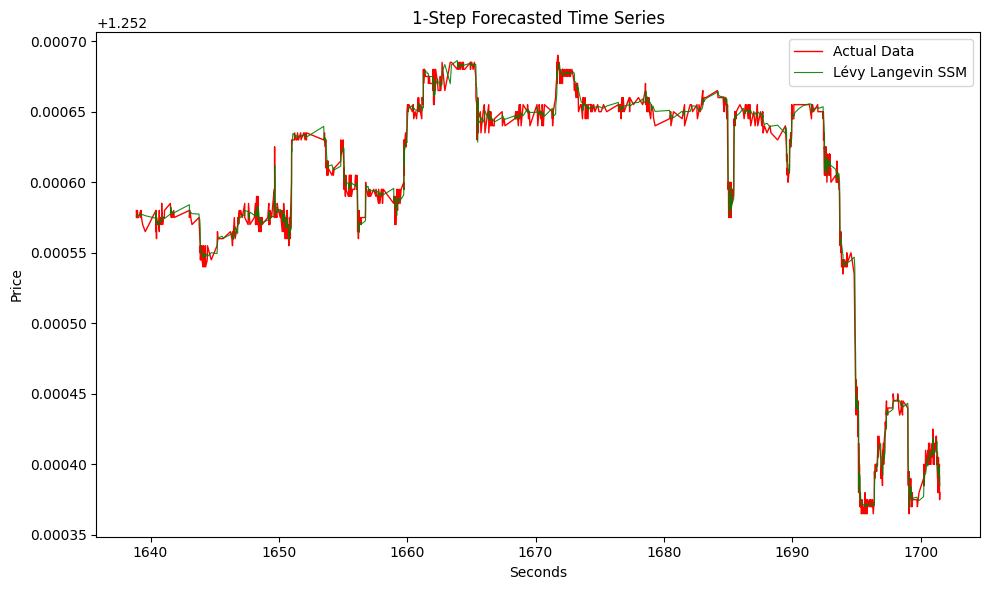}

    \end{subfigure}

    \caption{1-Step Forecasting of Gaussian and L\'evy Langevin Models on GBP/USD Data}
    \label{single step forecast comparison}
\end{figure}

The practical utility of our inference results is assessed by the forecasting performance. The testing dataset is obtained by offsetting the training data by $60,000$ ticks and taking the subsequent $30,000$ ticks. We compare our approach to the Gaussian Langevin model without subordination, which is trained on the same data using maximum likelihood via a Kalman filter \cite{KF_Reference} for all parameters. Forecasting is performed via the density $p(y_{t+n}|y_{1:t})$ and then takes the mean, which can be achieved by a Rao-Blackwellized particle filter and the Kalman filter for the 2 cases \cite{Kalman_Forecast,SMC_Forecast,particle_filters,Levy_State_Space_Model}. Details for forecasting in the $2$ models can be found in Supplement B \cite{SuppB}. An additional hit rate (directional accuracy) metric \cite{hit_rate_reference} is used to confirm that the models capture actual directional information. Figure \ref{single step forecast comparison} illustrates the single-step forecast performance for $1000$ ticks within the testing data, and it can be observed that the L\'evy model captures the FX data much better, especially when there is a jump.

Figure \ref{FX_Multi_Step_Forecast_MSE} and \ref{FX_Multi_Step_Forecast_Hit_Rates} show the multi-step forecasting performance in terms of MSE and hit rates. The L\'evy Langevin model learned has better MSEs across all steps, and considerable improvements have been introduced in forecasts over a few steps until the accumulated errors dominate. In terms of hit rates, the L\'evy model also outperforms the Gaussian case for all forecast horizons. These factors together demonstrate the practical utility of our inference results.

\begin{center}
  \begin{minipage}{0.45\textwidth}
    \centering
    \includegraphics[width=\textwidth]{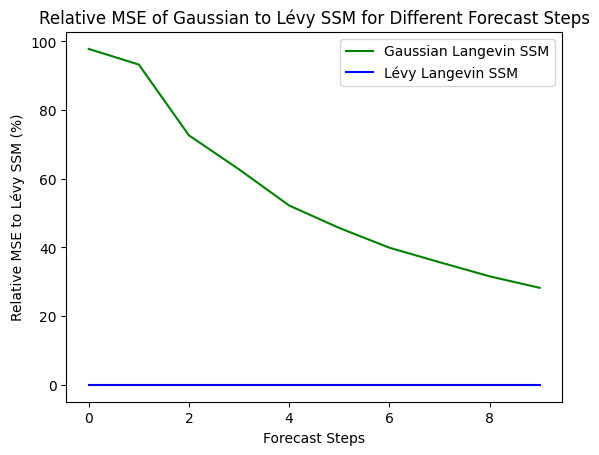}
    \captionof{figure}{Multi-Step Forecast MSE}
    \label{FX_Multi_Step_Forecast_MSE}
  \end{minipage}
  \hfill
  \begin{minipage}{0.45\textwidth}
    \centering
    \includegraphics[width=\textwidth]{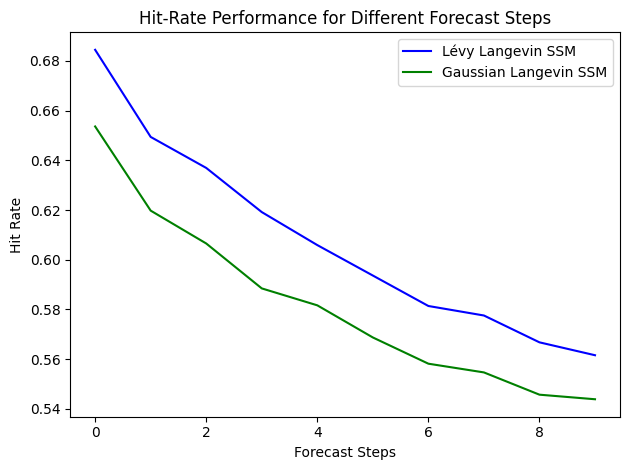}
    \captionof{figure}{Multi-Step Forecast Hit Rates}
    \label{FX_Multi_Step_Forecast_Hit_Rates}
  \end{minipage}
\end{center}

\section{Conclusions}
In this work, a novel Bayesian non-parametric framework has been proposed to address the challenging problem of inferring subordinator and NVM L\'evy measures under the L\'evy state space model by employing a flexible random measure prior, the Independent Gamma-scaled Dirichlet process. Various properties and conjugacy results about this random measure and its mixture model have been discussed. 
An effective MCMC algorithm has been developed to realize the inference, and efficiency and accuracy of our method are demonstrated via experimental results on both synthetic and tick-level financial data.
Future research directions could include extending the framework to non-linear system dynamics and multivariate L\'evy processes.

\section*{Code Availability Statement}
The code for our algorithms and experiments is available at: \url{https://github.com/zhl24/bayesian-np-inference-levy-measures-ssm.git}.

\begin{supplement}
\stitle{Supplement A}
\sdescription{Additional Preliminaries.}
\end{supplement}
\begin{supplement}
\stitle{Supplement B}
\sdescription{Supplementary Details for Analysis and Experiments. }
\end{supplement}

\bibliographystyle{ba} 
\bibliography{bnp_levy}

\newpage

\appendix
\section{Additional Preliminaries}
\subsection{Simulation Algorithms for Compound Poisson Processes}
\label{simulation algorithm appendix}

The following algorithm for the finite activity case, i.e. the compound Poisson process simulation, is from \cite{tankov_book} and serves as an important building block for our simulation framework.

\begin{algorithm}[H]
\caption{Compound Poisson Process Simulation}
\label{compound Poisson process simulation algorithm}
\begin{itemize}
    \item Simulate a random variable $N$ from Poisson distribution with parameter $\lambda T$. $N$ gives the total number of jumps in the interval $[0,T]$.
    \item Simulate $N$ independent random variables $U_i$ uniformly distributed in the interval $[0,T]$. These variables correspond to the jump times.
    \item Simulate jump sizes: $N$ independent random variables $Y_i$ with law $f(dx)=\frac{\nu(dx)}{\lambda}$.
\end{itemize}
\end{algorithm}

\subsection{Inference of the Conditional L\'evy State Space Model}
\label{Inference of the Conditional Levy State Space Model Appendix}

In this section, we show first the conditionally Gaussian structure in the subordinated SSM, and then present how inference may be achieved with marginalization to the NVM parameters in the NVM process in the partial and complete marginalization cases.

A key feature of the system response in equation (19) of the main article is its conditionally Gaussian nature on the subordinator series:

\begin{equation}
    \label{conditional distribution of the system response}
    X(t)|X(s),\{(Z_i,V_i)\}_{V_i \in(s,t]},\mu_w,\sigma_w^2 \sim N(e^{A(t-s)}X(s) +  m_{(s,t]}, C_{(s,t]}),
\end{equation} where $m_{(s,t]}$ is the mean vector , and $C_{(s,t]}$ is the covariance matrix of the summation term in the main article's equation (13). The specific forms are given below:

\begin{equation}
    \label{Conditional Gaussian mean}
    m_{(s,t]} =\mu_w\sum_{i:s< V_i\leq t}Z_ie^{A(t-V_i)}h = \mu_w \bar{m}_{(s,t]}
\end{equation}

\begin{equation}
    \label{Conditional Gaussian Covariance}
         C_{(s,t]} =\sigma_w^2 \sum_{i:s< V_i \leq t}Z_i  (e^{A(t-V_i)}h)(e^{A(t-V_i)}h)^T = \sigma_w^2 \bar{ C}_{(s,t]},
\end{equation} where $\bar{m}_{(s,t]}$ and $\bar{ C}_{(s,t]}$ are the normalized mean and covariance with respect to $\mu_w$ and $\sigma_w^2$, respectively. This structure allows a Kalman filter to be applied for the conditional inference.

To marginalize $\mu_w$, an extended state vector $\alpha$ that has $\mu_w$ as an additional state compared to $X(t)$ is used. By re-arranging the terms in equation (19) of the main article, the extended state vector and the corresponding conditional SDE, which is also Gaussian, are given below:

\begin{equation}
    \label{Extended State System Equation}
    \alpha (t)= 
    \begin{bmatrix}
        X(t) \\
        \mu_w \\
    \end{bmatrix}, \quad \alpha(t)=\hat{A}\alpha(s) + \hat{B}e_{(s,t]},\quad  e_{(s,t]}\sim N(0,C_{(s,t]}),
\end{equation}with

\begin{equation}
    \label{Extended State System Matrices}
    \hat{A}=\begin{bmatrix}
        e^{A(t-s)} & \bar{m}_{(s,t]} \\
        0 & 1 \\
\end{bmatrix}, \quad \hat{B} = \begin{bmatrix}
       I_{P\times P} \\
        0_{1 \times P}  \\
\end{bmatrix},
\end{equation} where $P$ is the dimension of the $P\times 1$ state vector $X(t)$. The emission model for the observation also needs to be modified to

\begin{equation}
    \label{emission structure for the extended state system}
    Y(t) = \hat{H} \alpha(t) + V(t),\quad V(t) \sim N(0,C_v),
\end{equation} with \begin{equation}
    \label{Extended State Emission Matrix}
    \hat{H}=
\begin{bmatrix}
        H,0_{M\times 1}
\end{bmatrix},
\end{equation} where $M$ is the dimension of the $M\times 1$ observation vector $Y(t)$. Equations \eqref{Extended State System Equation} and \eqref{emission structure for the extended state system} define the conditionally Gaussian model for the extended state $\alpha(t)$ that allows $\mu_w$ to be marginalized by running a Kalman filter on the extended state system directly, leading to the following predictive and filtering distributions conditional on the $\sigma_w^2$ parameter:

\begin{equation}
    \label{Kalman Predictive Distribution conditional on Sigmaw2}
    p(\alpha(t_n)|\{Y(t_i)\}_{i=1}^{n-1},\{(Z_i,V_i)\}_{V_i \leq t_n},\sigma_w^2) = N(\mu_{t_n|t_{n-1}},C_{t_n|t_{n-1}}),
\end{equation} 

\begin{equation}
    \label{Kalman Filtering Distribution in the Sigmaw2 conditional case}
    p(\alpha(t_n)|\{Y(t_i)\}_{i=1}^{n},\{(Z_i,V_i)\}_{V_i \leq t_n},\sigma_w^2) = N(\mu_{t_n|t_{n}},C_{t_n|t_n}).
\end{equation} Substituting the predictive distribution \eqref{Kalman Predictive Distribution conditional on Sigmaw2} into the emission model \eqref{emission structure for the extended state system}, we have the observation marginal conditional on $\sigma_w^2$:
\begin{equation}
    \label{observation marginal conditional on Sigmaw^2}
    p(Y(t_n)|\{Y(t_i)\}_{i=1}^{n-1},\{(Z_i,V_i)\}_{V_i \leq t_n},\sigma_w^2) = N(Y(t_n);\hat{H}\mu_{t_n|t_{n-1}},\hat{H}C_{t_n|t_{n-1}}\hat{H}^T + C_v). 
\end{equation} The log likelihood for the subordinator series conditional on $\sigma_w^2$ is therefore:

\begin{equation}
    \label{log likelihood conditional on Sigmaw^2}
    \begin{aligned}
        &\log p(\{Y(t_i)\}_{i=1}^N|\{(Z_i,V_i)\}_{V_i \leq t_N},\sigma_w^2)= - \frac{MN}{2}\log 2\pi - \frac{1}{2} \sum_{n=1}^N \log |\hat{H}C_{t_n|t_{n-1}}\hat{H}^T + C_v|\\
        &-\frac{1}{2}\sum_{n=1}^N (Y(t_n)-\hat{H}\mu_{t_n|t_{n-1}})^T (\hat{H}C_{t_n|t_{n-1}}\hat{H}^T + C_v)^{-1}(Y(t_n)-\hat{H}\mu_{t_n|t_{n-1}})\\
        &= - \frac{MN}{2}\log 2\pi - \frac{1}{2}\sum_{n=1}^N \log |F_{t_n}| - \frac{1}{2} \sum_{n=1}^N (Y(t_n) - \hat{Y}(t_n))^T  F_{t_n}^{-1} (Y(t_n) - \hat{Y}(t_n))\\
        & = - \frac{MN}{2}\log 2\pi - \frac{1}{2}\sum_{n=1}^N \log |F_{t_n}| - \frac{1}{2} E_N,
    \end{aligned}
\end{equation} where $M$ is the dimension of $Y$, $\hat{Y}(t_n)=\hat{H}\mu_{t_n|t_{n-1}}$, $F_{t_n} = \hat{H}C_{t_n|t_{n-1}}\hat{H}^T + C_v$, and $ E_N = \sum_{i=1}^N (Y(t_i)-\hat{Y}(t_i))^T F_{t_i}^{-1}(Y(t_i)-\hat{Y}(t_i))$.

Next, to marginalize $\sigma_w^2$, we need to define all the covariance matrices relative to $\sigma_w^2$. The normalized observation noise covariance is defined as $\bar{C}_v = \frac{1}{\sigma_w^2} C_v$ and the normalized dynamic noise covariance is $\bar{C}_{(s,t]}$ defined before in  \eqref{Conditional Gaussian Covariance}, which allows us to isolate the dependence of $\sigma_w^2$ in the system. Running a Kalman filter on the normalized system, we may re-apply the $\sigma_w^2$ factor to the covariances to obtain the filtering and incremental marginal distributions conditional on $\sigma_w^2$:

\begin{equation}
    \label{Hidden State Posterior conditional on sigmaw}
    p(\alpha(t_n)|\{Y(t_j)\}_{j=1}^n,\sigma_w^2,\{(Z_i,V_i)\}_{V_i \leq t_n}) = N(\mu_{t_n|t_n},\sigma_w^2 \bar{C}_{t_n|t_n}),
\end{equation}

\begin{equation}
    \label{single observation marginal conditional on sigmaw}
    p(Y(t_n)|\{Y(t_j)\}_{j=1}^{n-1},\sigma_w^2, \{(Z_i,V_i)\}_{V_i \leq t_n}) = N(\hat{Y}(t_n),\sigma_w^2 F_{t_n}),
\end{equation} where $\mu_{t_n|t_n}$ and $\bar{C}_{t_n|t_n}$ are the Kalman filter output variables for the normalized system, and re-define $F_{t_n} = \hat{H}\bar{C}_{t_n|t_{n-1}}\hat{H}^T + \bar{C}_v$. To marginalize $\sigma_w^2$ in the conditional distributions, we may introduce a conjugate inverse Gamma prior to $\sigma_w^2$ as $ p(\sigma_w^2) = \text{IG}(\alpha_w,\beta_w)$. This also leads to an Inverse Gamma posterior for $\sigma_w^2$ that allows inference of this parameter:

\begin{equation}
    \label{Inverse Gamma Conjugate Posterior}
    p(\sigma_w^2|\{Y(t_j)\}_{j=1}^N,\{(Z_i,V_i)\}_{V_i \leq t_N}) = \text{IG}(\alpha_w+\frac{MN}{2},\beta_w+\frac{E_N}{2}) ,
\end{equation} where $E_N = \sum_{i=1}^N (Y(t_i)-\hat{Y}(t_i))^T F_{t_i}^{-1}(Y(t_i)-\hat{Y}(t_i))$. By integrating the multivariate Gaussian with respect to the univariate Inverse Gamma posterior for $\sigma_w^2$, the marginalized filtering distribution can be shown to be a multivariate Student-t distribution,

\begin{equation}
    \label{marginalized posterior for extended state}
    p(\alpha(t_n)|\{Y(t_j)\}_{j=1}^n,\{(Z_i,V_i)\}_{V_i \leq t_n}) = t_{2(\alpha_w+\frac{Mn}{2})}(\mu_{t_n|t_n},\frac{\beta_w+\frac{E_n}{2}}{\alpha_w + \frac{Mn}{2}}\bar{C}_{t_n|t_n}),
\end{equation} and the log marginal for all observations is

\begin{equation}
    \begin{aligned}
    &\log\big(p(\{Y(t_j)\}_{j=1}^N|\{(Z_i,V_i)\}_{V_i \leq t_N})\big) = -\frac{MN}{2}\text{log}2\pi - \frac{1}{2}\sum_{i=1}^N \text{log} |F_{t_i}| + \alpha_w\text{log}\beta_w\\
    &-\text{log}\Gamma(\alpha_w)+\text{log}\Gamma(\alpha_w + \frac{MN}{2})-(\alpha_w+\frac{MN}{2})\text{log}(\beta_w+\frac{E_N}{2}).
    \end{aligned}
\end{equation} This fully marginalized scheme can be adjusted straightforwardly to the case with $\mu_w$ conditional and $\sigma_w^2$ marginalized by restoring the state vector to $X$.

Finally, we note that equations \eqref{Hidden State Posterior conditional on sigmaw} and \eqref{Inverse Gamma Conjugate Posterior} jointly define the posterior Normal-Inverse-Gamma distribution for $(\mu_w,\sigma_w^2)$:
\begin{equation}
    \label{final posterior distirbution for muw}
    \pi(\mu_w|\sigma_w^2) = N(\mu_w;\mu',\sigma_w^2 k_w'),
\end{equation}
\begin{equation}
    \label{final posterior distribution for sigmaw2}
    \pi(\sigma_w^2) = \text{IG}(\sigma_w^2;\alpha_w',\beta_w'),
\end{equation}where the posterior parameters are:
\begin{equation}
    \label{NIG Posterior Parameter Definitions}
    \begin{aligned}
        & \alpha_w' = \alpha_w + \frac{MN}{2}, \quad
\beta_w' = \beta_w + \frac{E_N}{2},\\
    & \mu' = e_{\mu_w}^\top \mu_{t_N|t_N}, \quad
k_w' = e_{\mu_w}^\top \bar{C}_{t_N|t_N}e_{\mu_w},
    \end{aligned}
\end{equation} where $N$ is the observation length, and $\mu_{t_n|t_m}$ and $\bar{C}_{t_n|t_m}$ denote the standard Kalman filter outputs conditional on $\sigma_w^2$. 
The vector $e_{\mu_w}$ is the state selection vector corresponding to the augmented state $\mu_w$.

\subsection{Finite DP Sampler}
\label{Finite DP Appendix}
This section contains the details of a finite DP or blocked Gibbs sampler in \cite{Ishwaran_James_2001} as an example of the conditional sampler for DP, which has also been used in our experiments. Detailed instructions have been given by \cite{Ishwaran_James_2001} on how to select an appropriate truncation level, together with discussions about other properties of the truncated form and its mixing efficiency. The specific steps we use to draw from a posterior DP \eqref{DP Posterior Formula} are given below:

\begin{enumerate}
    \item Draw samples $\{\beta_i\}_{i=1}^{K-1}$ from the Beta distribution $Beta(1,\alpha+n)$, and set $\beta_K=1$ for normalization. Then compute the probability weights $\{w_i\}_{i=1}^K$ via:

    \begin{equation}
        \label{Dirichelt Weights Computation}
        w_k = \beta_k \prod_{i=1}^{k-1}(1-\beta_i).
    \end{equation}

    \item Draw the discrete measure locations $\{x_i\}_{i=1}^K$ iid from the posterior base distribution in \eqref{DP Posterior Formula}.
    
\end{enumerate} The sample random measure drawn from the posterior Dirichlet process then has the following finite summation representation:

\begin{equation}
    \label{Truncated Posterior DP Sample}
    G(\cdot) = \sum_{i=1}^K w_i \delta_{x_i}(\cdot),
\end{equation} which is essentially a finite discrete distribution.

\subsection{DP Hyper-Parameter Sampling}
\label{Appendix: DP Hyper-parameter sampling}
This section provides more details about the DP hyper-parameter sampling scheme used in our experiments. 

Recall that for the DP model complexity parameter $\alpha$, its likelihood has been shown in \cite{DP_alpha_likelihood} to be:

\begin{equation}
    p(k|\alpha,n) \propto \alpha^k \frac{\Gamma(\alpha)}{\Gamma(\alpha+n)}.
\end{equation} By introducing $p(\alpha)\sim Gamma(a,b)$, where $a$ and $b$ are the shape and rate parameters respectively, \cite{DP_Alpha_Gibbs_Sampler} constructs a Gibbs sampler for sampling $\alpha$ by introducing an auxiliary variable $\phi$ with $Beta(\phi;\alpha+1,n)$ distribution, leading to a mixture of Gamma distribution for $\alpha$:

\begin{equation}
    \label{mixture of Gamma for alpha posterior}
    p(\alpha|\phi,k,n) = \rho \, Gamma(a+k, b-\log \phi) + (1-\rho) \, Gamma(a+k-1,b-\log \phi),
\end{equation} with mixing ratio:

\begin{equation}
    \label{mixing ratio}
    \frac{\rho}{1-\rho} = \frac{a+k-1}{n(b-\log \phi)}.
\end{equation} This scheme can be further generalized by using a mixture of Gamma priors for $\alpha$. See \cite{DP_hyperparameter_tutorial} for more details. 

\newpage
\section{Supplementary Details for Analysis and Experiments}
\subsection{Complete Randomness of the IGSDP}

\begin{proposition}
\label{proposition: IGSDP not CRM in general}
    Let $\hat{G} \sim \text{IGSDP}(\alpha_\lambda,\beta_\lambda,\alpha,H)$. $\hat{G}$ is not completely random unless $\alpha_\lambda = \alpha$, where $\hat{G}$ reduces to a Gamma process. 
\end{proposition}

\begin{proof}
    Because we know that the Gamma process is completely random, the claim follows from a simple covariance check \cite{first_course_in_prob} between $\hat{G}(A_1)$ and  $\hat{G}(A_2)$, where $A_1$ and $A_2$ are disjoint:

    \begin{equation}
        \label{Covariance Check for IGSDP being not a CRM}
        \begin{aligned}
            \mathrm{Cov}[\hat{G}(A_1),\hat{G}(A_2)] & = \mathbb{E}[\hat{G}(A_1) \hat{G}(A_2)] - \mathbb{E}[\hat{G}(A_1)] \mathbb{E}[\hat{G}(A_2)]\\
            & = \mathbb{E}[\lambda^2] \mathrm{Cov}[G(A_1),G(A_2)] + \mathrm{Var}[\lambda]H(A_1)H(A_2)\\
            & = H(A_1)H(A_2) \frac{\alpha_\lambda (\alpha - \alpha_\lambda)}{(\alpha+1)\beta_\lambda^2},
        \end{aligned}
    \end{equation} where the last line is obtained from the covariance of the Dirichlet distribution of the Dirichlet process and the Gamma distribution for $\lambda$. It is then obvious that for the covariance to be $0$, we need $\alpha_\lambda = \alpha$, which corresponds to the Gamma process, and for other general IGSDP, independence does not hold. 
\end{proof}

\subsection{Projection Estimator Baselines}

In this section, we provide the implementation details for the $2$ projection estimators in \cite{figueroa-lopez_nonparametric_2004,figueroa2009nonparametric}. Given the discrete observations  $\{X_k\}_{k=1}^n$ of a L\'evy process $\{X(t)\}_{t\in [0,T]}$ at equally spaced time points, the jump sizes are approximated by the increments:
\begin{equation}
    \label{Jump Size Recovery in Projection Estimators}
    x_i = X_{i+1}-X_{i}, \quad i=1,\ldots,n-1.
\end{equation} 
For the projection estimators, the set of $D$-dimensional orthogonal basis functions $\{\varphi_d\}_{d=1}^D$ we study here is a collection of normalized indicator functions, which are $\varphi_d(x) = \frac{1}{\sqrt{x_d-x_{d-1}}}\mathbb{I}(x_{d-1}\leq x < x_d)$. The projection estimator in the Lebesgue domain is then:
\begin{equation}
    \label{Lebesgue projection Estimator Definition}
    \hat{Q}(x) = \sum_{d=1}^D \beta_d \varphi_d(x),
\end{equation} 
where $\beta_d$ are the projections estimated by:
\begin{equation}
    \label{Lebesgue Projection Estimator Projections}
    \beta_d = \frac{1}{T}\sum_{i=1}^{n-1} \varphi_d(x_i).
\end{equation} 

The regularized projection estimator is constructed by writing \(Q(x)=g(x)s(x)\), where the regularization function is chosen as \(g(x)=x^{-2}\), as in \cite{figueroa-lopez_nonparametric_2004}. The estimator is
\begin{equation}
    \label{Regularized projection Estimator}
    \hat{Q}_{\mathrm{reg}}(x)
    =
    g(x)\sum_{d=1}^D\alpha_d\varphi_d(x)
    =
    g(x)\sum_{d=1}^D
    \left(\frac{1}{T}\sum_{i=1}^{n-1}\frac{\varphi_d(x_i)}{g(x_i)}\right)\varphi_d(x).
\end{equation} 
In the NVM process experiments, we consider uniformly spaced normalized indicator basis functions, and the basis resolution $D$ is tuned by minimizing the total variation norm.

\subsection{Functional Autocorrelation Time}
\label{Functional Autocorrelation Time Derivation}

In this section, the details about the functional autocorrelation time used in the mixing performance evaluation are provided. According to \cite{MCMC_in_statistical_mechanics}, the definition of the autocorrelation function is:

\begin{equation}
    \label{autocorrelation time definition}
    \rho(t) = \frac{C(t)}{C(0)},
\end{equation} where $t$ is the lag in the iteration steps of the Markov chain. The main problem is in the definition of $C(t)$. In the 1-dimensional problems, $C(t)$ is just the centered expected value of the product of samples at distance $t$, as shown below:

\begin{equation}
    \label{C in scalar case}
    C(t) = \mathbb{E}[(\theta_s-\mathbb{E}[\theta]) (\theta_{s+t}-\mathbb{E}[\theta])].
\end{equation} However, when the samples become functions, the definition of the product becomes ambiguous. Here, since we want a single scalar metric that measures the autocorrelation, we use the inner product definition:

\begin{equation}
    \label{inner product definition of the autocorrelation function}
    C(t) = \mathbb{E}[\int (f_s(x)-\mathbb{E}[f(x)]) (f_{s+t}(x)-\mathbb{E}[f(x)])dx]. 
\end{equation} To simplify the computation, we consider evaluating the sample functions on the same discrete axis $\{x_i\}_{i=1}^K$, and hence we may use a simple function or piecewise constant approximation for function $f(x)$: 

\begin{equation}
    \label{simple function approximation}
    f(x) \approx  \sum_{i=1}^{K-1}f(x_i) \mathbb{I}(x_i \leq x < x_{i+1}).
\end{equation} Substituting \eqref{simple function approximation} into \eqref{inner product definition of the autocorrelation function}, we have the following Monte Carlo estimator:

\begin{equation}
    \label{functional C estimator formula}
    C(t)\approx \sum_{i=1}^{K-1} \frac{x_{i+1}-x_i}{N-t}\sum_{s=1}^{N-t}(f_s(x_i)-\mathbb{E}[f(x_i)])(f_{s+t}(x_i)-\mathbb{E}[f(x_i)]),
\end{equation}where $f_s(x_i)$ is the $s$th sample of the function value at $x_i$, and the mean function $\mathbb{E}[f(x)]$ estimate is:

\begin{equation}
    \label{simple function mean estimate}
    \mathbb{E}[f(x)] \approx  \sum_{i=1}^{K-1} \frac{1}{N}\sum_{s=1}^N f_s(x_i) \mathbb{I}(x_i \leq x < x_{i+1}).
\end{equation} Equation \eqref{simple function mean estimate} can then be used as a plug-in estimator to \eqref{functional C estimator formula} to obtain an estimate for the functional autocorrelation function. The integrated autocorrelation time is then computed from the estimated autocorrelation function.

\subsection{Ground Truth Generation for NVM process L\'evy Density}

\label{nvm ground truth generation appendix}

In this section, we describe how we generate the NVM L\'evy density ground truth. For most NVM densities, they do not have a closed-form solution, and therefore, numerical integration has to be done. We make use again of the fact that for subordinator L\'evy processes without closed-form likelihood, explicit truncation has to be applied \cite{tankov_book}. We then consider the NVM L\'evy measure corresponding to such truncated subordinator:

\begin{equation}
    \label{Approximate NVM Density Formula}
    \hat{Q}_{NVM}(x) = \int_{\epsilon}^\infty N(x;\mu_wz,\sigma_w^2 z) Q(z)dz,
\end{equation} where $\epsilon > 0$ is some deterministic truncation threshold on the subordinator. Knowing that the truncated process is of finite activity, we may re-write the integral as:

\begin{equation}
    \label{Approximate NVM Density Formula Evaluation}
    \hat{Q}_{NVM}(x) = \int_{\epsilon}^\infty N(x;\mu_wz,\sigma_w^2 z) \lambda f(z) dz,
\end{equation} where $\lambda = \int_{\epsilon}^\infty Q(z)dz$ is the overall activity in the truncated range and the function, $f(x), x\in [\epsilon,\infty)$, is the corresponding jump size distribution. Since $f(z)$ is a valid probability density function, and the integration range is also the support of $f(z)$, we may estimate the approximate NVM density via a Monte-Carlo estimator as shown below:

\begin{equation}
    \label{Monte Carlo Estimator for the NVM Density}
    \hat{Q}_{NVM}(x) \approx \lambda \frac{1}{N}\sum_{i=1}^N N(x;\mu_w z_i,\sigma_w^2z_i),
\end{equation} with $\{z_i\}_i$ being the jump sizes drawn from the distribution $f(x)$. Another difficulty in this estimator is in the evaluation of the truncated activity $\lambda$, but since we are sure that it is a finite integral, numerical integration techniques may be used. Therefore, using \eqref{Monte Carlo Estimator for the NVM Density}, the ground truth for the NVM process density may be estimated by generating a large number of jumps from the truncated distribution, using algorithms, for instance, in \cite{tankov_book,Yaman_GH_Simulation,Yaman_gig}. 

\subsection{Marginal Sample Recovery in the Collapsed Gibbs Scheme}

In this section, we explain how marginalized variables in the MCMC sampler can be recovered for posterior inference. The stationary distribution of the collapsed Gibbs sampler in our case is
\[
p(Q,\alpha,\{(Z_i,V_i)\}_{V_i \leq t_N},\theta|\{Y_j\}_{j=1}^N),
\]
where variables such as \(\sigma_w^2\), \(\{X_i\}\), and \(\{J_i\}\) have been analytically marginalized. This reduces the state space of the Markov chain, which can improve mixing and reduce Monte Carlo variance in posterior estimates \cite{collapsed_Gibbs_liu1994collapsed}. 

In general, consider a joint posterior distribution \(p(A,C|B)\). A collapsed scheme can be applied if
\[
p(A|B)=\int p(A,C|B)\,dC
\]
is analytically tractable. If the collapsed chain produces samples \(A^{(m)}\sim p(A|B)\), then posterior inference for the marginalized variable \(C\) can be recovered via
\begin{equation}
    \label{general collapsed state recovery scheme}
    \begin{aligned}
        p(C|B) 
        &= \int p(C|A,B) p(A|B)\,dA\\
        & \approx \frac{1}{M}\sum_{m=1}^M p(C|A^{(m)},B),
    \end{aligned}
\end{equation}
where \(\{A^{(m)}\}_{m=1}^M\) are samples from the collapsed posterior \(p(A|B)\). Drawing \(C^{(m)}\sim p(C|A^{(m)},B)\) then gives marginal samples from \(p(C|B)\).

In our setting,
\[
A=(Q,\alpha,\{(Z_i,V_i)\}_{V_i \leq t_N},\theta),
\qquad
B=\{Y_j\}_{j=1}^N.
\]
The marginalized variables, such as \(\sigma_w^2\), \(\{X_i\}\), and \(\{J_i\}\), are recovered by drawing from their corresponding conditional posterior distributions given each collapsed MCMC sample.

\newpage
\subsection{Additional Training Results with FX Data}

\begin{figure}[H]
    \centering
    \includegraphics[width=0.6\linewidth]{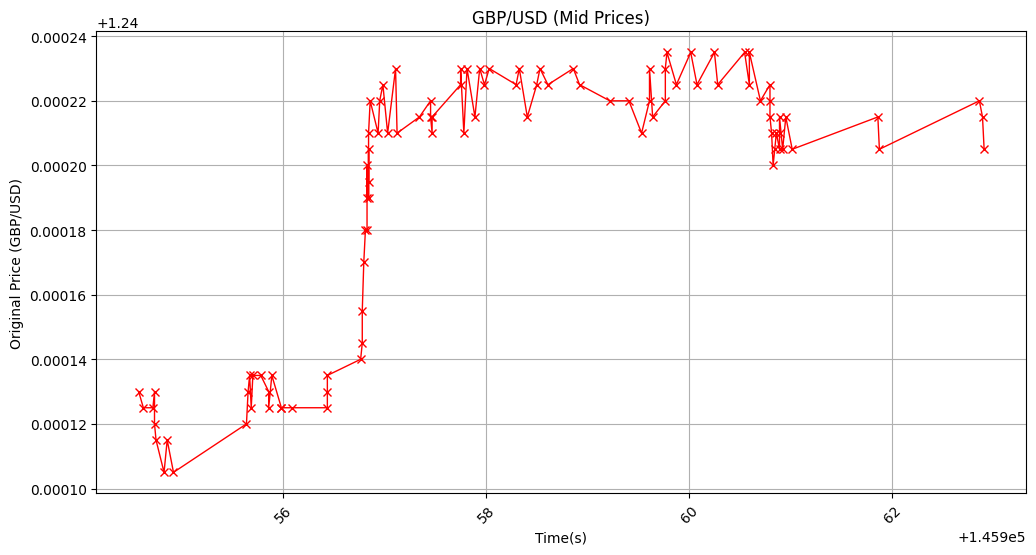}
    \caption{GBP/USD Data}
    \label{FX_tick_data}
\end{figure}

\begin{figure}[H]
    \centering
    \begin{subfigure}{0.4\textwidth}
        \centering
        \includegraphics[width=\textwidth]{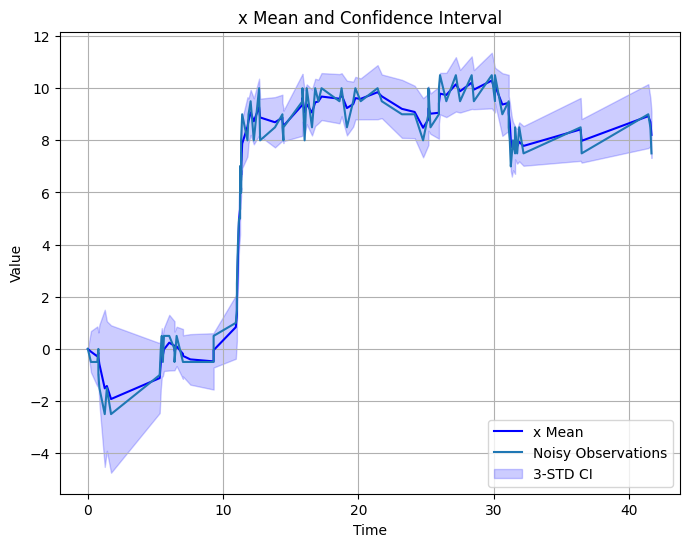}
        \label{FX_x_inference}
    \end{subfigure}
    \hfill
    \begin{subfigure}{0.4\textwidth}
        \centering
        \includegraphics[width=\textwidth]{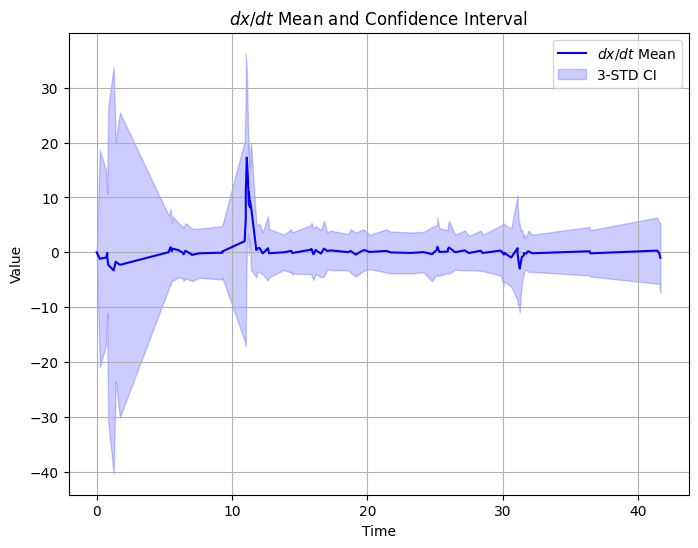} 
        \label{FX_dx_dt_inference}
    \end{subfigure}

    \begin{subfigure}{0.4\textwidth}
        \centering
        \includegraphics[width=\textwidth]{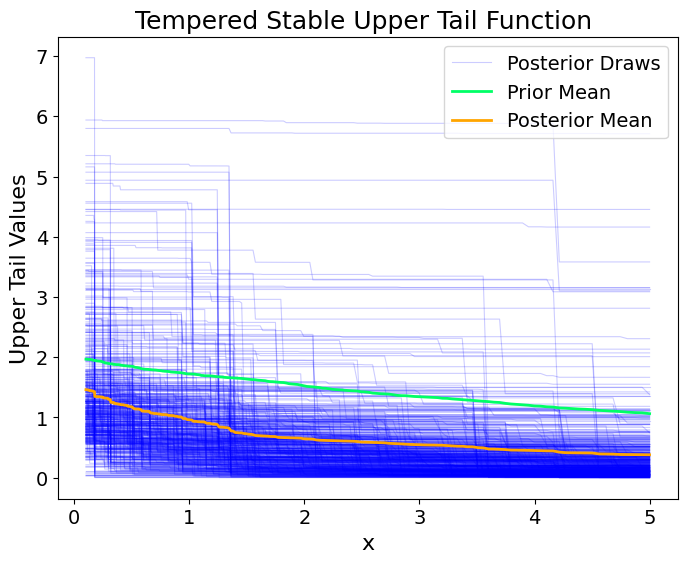}
\label{FX_subordinator_measure_inference}
    \end{subfigure}
    \hfill
    \begin{subfigure}{0.4\textwidth}
        \centering
        \includegraphics[width=\textwidth]{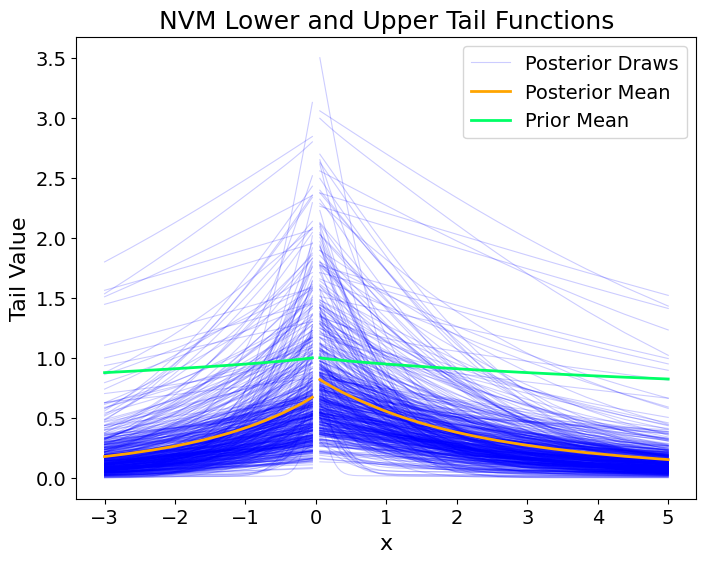}
    \end{subfigure}
    
    \caption{Inference Results of the GBP/USD Data}
    \label{FX_inference_results}
\end{figure}
In this section, we provide the details for the training of the Langevin model on FX data. We train our model on the FX data in figure \ref{FX_tick_data}, while also extracting the momentum signal \cite{levy_momentum_trading}. First, we normalize the data by selecting appropriate units based on the order of magnitude of the variations. For the initial subordinator series, we sample from a tempered stable process with parameters chosen to provide comparable magnitudes to the normalized data. Then, we run our algorithm for $250,000$ iterations, with the first $50,000$ iterations discarded as burn-in. For the prior model, we put $\alpha\sim\Gamma(1,1)$, $Q(z)|\alpha \sim \text{IGSDP}(2,1,\alpha,\Gamma(1,\frac{1}{8}))$, and $\mu_w\sim N(0,100)$. 

\begin{figure}[H]
    \centering
    \includegraphics[width=0.5\linewidth]{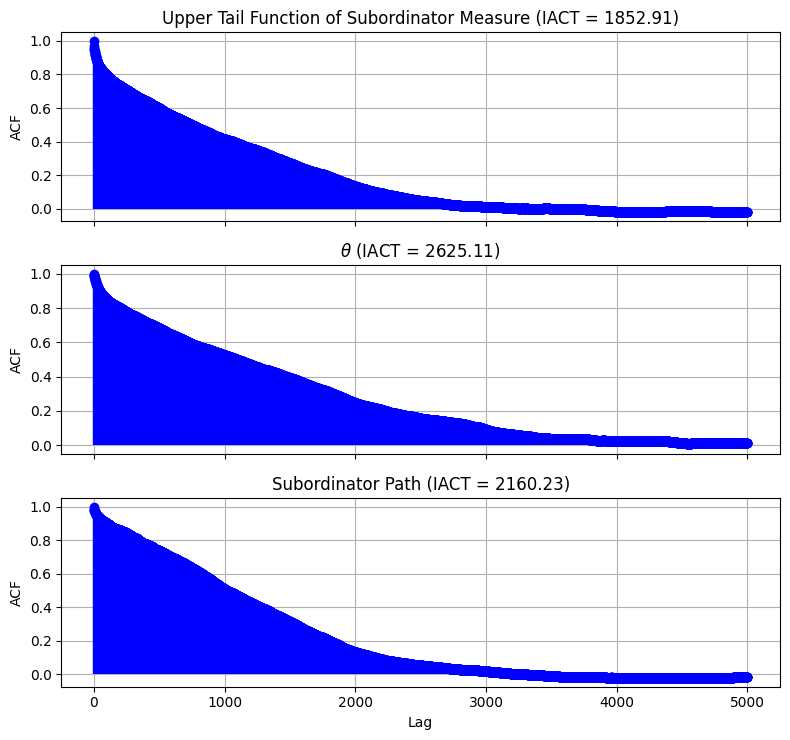}
    \caption{Mixing Performance of GBP/USD Data Inference}
    \label{FX_mixing_performances}
\end{figure}

Figure \ref{FX_inference_results} and \ref{FX_mixing_performances} present the main inference results and the mixing performance. Figure \ref{FX_mixing_performances} shows comparable auto-correlation time to the synthetic data case, which proves effective mixing, validating the inference results. Figure \ref{FX_inference_results} shows that the inferred hidden states successfully capture variations and large jumps in the data, and the momentum (velocity) signal exhibits a strong response during a large upward price jump and decays as the jump stops, aligning with expected market dynamics. These results serve as strong evidence for the validity of the L\'evy measures inferred.

\subsection{Forecasting in Gaussian Langevin Model}
\label{Gaussian Langevin Model Appendix}
This section includes the details to perform forecasting in the Gaussian Langevin model. For the Langevin model, we have the matrix exponential:

\begin{equation}
    \label{Langevin expm formula}
    \exp(A
    \Delta t)= \text{exp}(\theta \Delta t)\begin{bmatrix}
        0 & \frac{1}{\theta} \\
        0 & 1 \\ 
\end{bmatrix} + \begin{bmatrix}
        1 & -\frac{1}{\theta} \\
        0 & 0 \\ 
\end{bmatrix}.
\end{equation} A linear Gaussian SDE has a closed-form solution for discretization that is:

\begin{equation}
    \label{Discretization of Linear Gaussian SDE}
    x(t+\Delta t) = \exp(A\Delta t)x(t) + \epsilon_t, \quad \epsilon_t \sim N(0,C),
\end{equation} where the transition noise covariance matrix $C$ is\begin{equation}
    \label{Gaussian Transition Noise Covariance in Linear SDE Discretization}
    C = \sigma^2 \int_0^{\Delta t} \exp(A(\Delta t - \tau))M M^T \exp(A^T(\Delta t-\tau)) d\tau.
\end{equation} In the Gaussian Langevin case, the integral has a closed-form solution to be:

\begin{equation}
    \label{Langevin Transition Noise Covariance}
    C = \begin{bmatrix}
        \frac{\sigma^2}{\theta^2}(\frac{3}{2\theta}+\Delta t-2\frac{\exp(\theta\Delta t)}{\theta} + \frac{\exp(2\theta \Delta t)}{2\theta}) & \frac{\sigma^2}{\theta^2}(\frac{1}{2}+\frac{\exp(2\theta\Delta t)}{2}-\exp(\theta\Delta t)) \\
        \frac{\sigma^2}{\theta^2}(\frac{1}{2}+\frac{\exp(2\theta \Delta t)}{2}-\exp(\theta \Delta t)) & \frac{\sigma^2}{2\theta}(\exp(2\theta \Delta t)-1) \\
\end{bmatrix}.
\end{equation} These results then allow the standard Kalman filter forecasting scheme \cite{Kalman_Forecast} to be applied.

\subsection{Forecasting in L\'evy State-Space Models}
\label{Levy Langevin Model Appendix}

This section presents the details for forecasting in the general linear state-space models driven by NVM processes or other processes of similar structures using a bootstrap Rao-Blackwellized particle filter \cite{particle_filters,Levy_State_Space_Model}. The main difficulty arises from the non-linear states which are both non-Gaussian and non-Markovian. For notational convenience, we denote $x_{n:m}:=\{X(t_i)\}_{i=n}^m$, $y_{n:m}:=\{Y(t_i)\}_{i=n}^m$, $\{(Z_i,V_i)\}_{n:m}:= \{(Z_i,V_i)\}_{V_i \in [t_{n-1},t_m]}$. After a filtering run, the outputs of the Rao-Blackwellized particle filter with $N$ particles are:

\begin{equation}
    \label{SMC approximate on the subordinator series posterior}
    \begin{aligned}
        p(\{(Z_i,V_i)\}_{1:T}|y_{1:T}) \approx \sum_{l=1}^N w_T^{(l)} \delta_{\{(Z_i,V_i)\}_{1:T}^{(l)}}(\{(Z_i,V_i)\}_{1:T}),
    \end{aligned}
\end{equation} where the whole trajectory is needed for the non-Markovian states, and the linear states have the filtering distribution:

\begin{equation}
    \label{Linear State Filtering Density output from the SMC}
    p(x_T|y_{1:T},\{(Z_i,V_i)\}_{1:T}^{(l)}) = t_{2(\alpha_w+\frac{T}{2})}(\mu_{T|T}^{(l)},\frac{\beta_w+\frac{E_T^{(l)}}{2}}{\alpha_w + \frac{T}{2}}\bar{C}_{T|T}^{(l)}) \quad  \forall l,
\end{equation} where $\mu_{T|T}$ and $\bar{C}_{T|T}$ are the marginalized Kalman filter outputs, and $\alpha_w$ and $\beta_w$ are the prior parameters to $\sigma_w^2$, see supplement A \cite{SuppA} for more details about the filtering scheme. If we condition on $\sigma_w^2$, the distribution is a Gaussian:

\begin{equation}
    \label{Conditional Distribution for the filteirng density}
     p(x_T|y_{1:T},\{(Z_i,V_i)\}_{1:T}^{(l)},\sigma_w^2) = N(\mu_{T|T}^{(l)},\sigma_w^2 \bar{C}_{T|T}^{(l)})\quad \forall l.
\end{equation}

For forecasting, the first step would be forecasting the subordinator series by $n$ time frames, which can be approximated by the empirical distribution via:

\begin{equation}
    \label{subordinator series forecasting}
    \begin{aligned}
        &p(\{(Z_i,V_i)\}_{1:T+n}|y_{1:T})  = p(\{(Z_i,V_i)\}_{1:T}|y_{1:T}) \times p(\{(Z_i,V_i)\}_{T+1:T+n})\\
        &\approx \sum_{l=1}^N w_T^{(l)} \delta_{\{(Z_i,V_i)\}_{1:T}^{(l)}}(\{(Z_i,V_i)\}_{1:T}) \times p(\{(Z_i,V_i)\}_{T+1:T+n})\\
        &\approx \sum_{l=1}^N w_T^{(l)} \delta_{\{(Z_i,V_i)\}_{1:T+n}^{(l)}}(\{(Z_i,V_i)\}_{1:T+n}),
    \end{aligned}
\end{equation} where the last line is just using additional sampling, and note also that the particle weights at the final filtering step $T$ are inherited by the forecasting trajectories.

Now, we have all the tools to derive the forecasting distribution estimate. Because of the conditionally Gaussian structure, we start with the same step as the Kalman filter case, by integrating over the forecasted linear hidden state. We then marginalize over all the previous subordinator series and also the intermediate linear latent states $x_{T:T+n-1}$ to reveal the model structure:

\begin{equation}
    \label{Forecasting Distribution Derivation}
    \begin{aligned}
        &p(y_{T+n}|y_{1:T})  = \int p(y_{T+n},x_{T+n}|y_{1:T}) dx_{T+n}\\
        & = \iiint p(y_{T+n},x_{T+n},x_{T:T+n-1},\{(Z_i,V_i)\}_{1:T+n}|y_{1:T})\\
        & \quad \quad \quad \quad \quad \quad \quad \quad \quad dx_{T:T+n-1}dx_{T+n} d\{(Z_i,V_i)\}_{1:T+n},
    \end{aligned}
\end{equation}and note that this is very similar to the linear Gaussian state space model scenario but with an additional marginalization over the non-linear states:

\begin{equation}
    \label{Derivation for Forecasting in Rao Blackwellized PF section 2}
    \begin{aligned}
        p(y_{T+n}|y_{1:T}) &= \iiint p(\{(Z_i,V_i)\}_{1:T+n}|y_{1:T})p(x_T|y_{1:T},\{(Z_i,V_i)\}_{1:T}) \\
        & \times \prod_{t=1}^n p(x_{T+t}|x_{T+t-1},\{(Z_i,V_i)\}_{T+t}) g(y_{T+n}|x_{T+n})\\&dx_{T+n} dx_{T:T+n-1} d\{(Z_i,V_i)\}_{1:T+n}
    \end{aligned}
\end{equation} This form already allows for a completely simulation-based Monte Carlo inference scheme, but we can improve it by considering Rao-Blackwellization for variance reduction.

To obtain the Rao-Blackwellized inference scheme, we substitute the empirical distribution in \eqref{subordinator series forecasting} into \eqref{Derivation for Forecasting in Rao Blackwellized PF section 2} to obtain the weighted sum of conditionally Gaussian model forecasting results:

\begin{equation}
    \begin{aligned}
        &p(y_{T+n}|y_{1:T})\\
        &\approx  \sum_{l=1}^N w_T^{(l)} \iint p(x_T|y_{1:T},\{(Z_i,V_i)\}_{1:T}^{(l)}) \prod_{t=1}^n p(x_{T+t}|x_{T+t-1},\{(Z_i,V_i)\}_{T+t}^{(l)}) \\
        &\quad \quad \quad \quad \quad \quad \quad \quad\quad \quad \quad \quad\quad \quad \quad \quad \times g(y_{T+n}|x_{T+n})dx_{T+n} dx_{T:T+n-1}\\
        &=\sum_{l=1}^N w_T^{(l)}\iint p(x_T|y_{1:T},\{(Z_i,V_i)\}_{1:T}^{(l)}) \prod_{t=1}^n p(x_{T+t}|x_{T+t-1},\{(Z_i,V_i)\}_{T+t}^{(l)})  \\
        & \quad \quad \quad \quad \quad \quad \quad \quad\quad \quad \quad \quad\quad \quad \quad \quad  \times dx_{T:T+n-1} g(y_{T+n}|x_{T+n})dx_{T+n}\\
        &= \sum_{l=1}^N w_T^{(l)} \int p(x_{T+n}|y_{1:T},\{(Z_i,V_i)\}_{1:T+n}^{(l)}) g(y_{T+n}|x_{T+n}) dx_{T+n}.
    \end{aligned}
\end{equation} The integral can then be solved by the standard Kalman filter forecasting scheme conditional on $\sigma_w^2$. Furthermore, we may also marginalize over the NVM parameter $\sigma_w^2$, which requires an additional integration. The inner integral can then be solved by the Kalman scheme, and the outer integral turns the final distribution into a Student-t:

\begin{equation}
    \begin{aligned}
        &p(y_{T+n}|y_{1:T})\\
        &\approx \sum_{l=1}^N w_T^{(l)} \iint p(x_{T+n}|y_{1:T},\{(Z_i,V_i)\}_{1:T+n}^{(l)},\sigma_w^2) g(y_{T+n}|x_{T+n}) dx_{T+n} \\
        & \quad \quad \quad \quad \quad \quad \quad \quad \quad \quad \quad \quad \quad \quad \times p(\sigma_w^2|y_{1:T},\{(Z_i,V_i)\}_{1:T+n}^{(l)}) d\sigma_w^2 \\
        & = \sum_{l=1}^N w_T^{(l)} \iint N(x_{T+n};\mu_{T+n|T}^{(l)},\sigma_w^2 \bar{C}_{T+n|T}^{(l)}) N(y_{T+n};Bx_{T+n},\sigma_w^2 \bar{C}_v)dx_{T+n}  \\
        & \quad \quad \quad \quad \quad \quad \quad \quad \quad \quad \quad \quad \quad \quad   \times p(\sigma_w^2|y_{1:T},\{(Z_i,V_i)\}_{1:T+n}^{(l)}) d\sigma_w^2\\
        & = \sum_{l=1}^N w_T^{(l)} \int N(y_{T+n};B\mu_{T+n|T}^{(l)},\sigma_w^2(B\bar{C}_{T+n|T}^{(l)}B^T + \bar{C}_v))\\
        & \quad \quad \quad \quad \quad \quad \quad \quad \quad \quad \quad \quad \quad \quad \times IG(\sigma_w^2;\alpha_w+\frac{T}{2},\beta_w + \frac{E_T}{2}) d\sigma_w^2\\
        & = \sum_{l=1}^N w_T^{(l)} St(B\mu_{T+n|T}^{(l)},\sqrt{\frac{(\beta_w + \frac{E_T^{(l)}}{2})(B\bar{C}_{T+n|T}^{(l)}B^T + \bar{C}_v)}{\alpha_w + \frac{T}{2}}},2 \alpha_w + T).
    \end{aligned}
\end{equation} This leads to the mean prediction formula in our experiments:

\begin{equation}
    \label{Mean Estimate for the Forecast in the Rao-Blackwellized Particle Filter Case}
    \mathbb{E}[y_{T+n}|y_{1:T}] = \sum_{l=1}^N w_T^{(l)} B \mu_{T+n|T}^{(l)}.
\end{equation}Note that the mixture of Student-t distributions allows other statistics to be calculated too.

\end{document}